\documentclass[letterpaper,twocolumn,10pt]{article}
\usepackage{usenix,epsfig,endnotes}
\usepackage{url}

\usepackage{compactbib}
\usepackage[compact]{titlesec}
\renewcommand{\middle}{}
\usepackage[utf8x]{inputenc}

\usepackage[tight,footnotesize]{subfigure}

\usepackage{amsmath,amssymb,graphicx,bbm}
\usepackage[english]{babel}

\newcommand{\Epsilon}{\mathrm{E}}
\newcommand{\mathd}{\mathrm{d}}
\newcommand{\tmop}[1]{\ensuremath{\operatorname{#1}}}
\newcommand{\tmsamp}[1]{\textsf{#1}}
\newcommand{\tmstrong}[1]{\textbf{#1}}

\newcommand{\tmtexttt}[1]{{\ttfamily{#1}}}

\author{
{\rm Pili Hu$^*$, Qijiang Fan$^+$ and Wing Cheong Lau$^*$} \\
{\small $^*$ Department of Information Engineering, The Chinese University of Hong Kong }\\
{\small $^+$ School of Computer Science and Technology, Huazhong University of Science and Technology}
} %

\title{
SNSAPI: A Cross-Platform Middleware for Rapid Deployment of Decentralized Social Networks
}

\begin{document}

\maketitle

\begin{abstract}
In this paper, we present the design, implementation and our year-long maintenance experience of
SNSAPI, a Python-based middleware which unifies the interfaces and data structures of
heterogeneous Social Networking Services (SNS). Unlike most prior works,
our middleware is user-oriented and requires zero infrastructure support.
It enables a user to readily conduct online social activities in a programmable, cross-platform fashion while
gradually reducing the dependence on centralized Online Social Networks (OSN).
More importantly, as the SNSAPI middleware can be used to support decentralized social networking services
via conventional communication channels such as RSS or Email, it enables the deployment of
Decentralized Social Networks (DSN) in an incremental, ad hoc manner.
To demonstrate the viability of such type of
DSNs, we have deployed an experimental 6000-node SNSAPI-based DSN on PlanetLab and evaluate its performance
by replaying traces of online social activities collected from a mainstream OSN.
Our results show that, with only mild resource consumption, the SNSAPI-based DSN can achieve acceptable forwarding latency
comparable to that of a centralized OSN.
We also develop an analytical model to
characterize the trade-offs between resource consumption and message forwarding
delay in our DSN. Via 20 parameterized experiments on PlanetLab, we have
found that the empirical measurement results match reasonably with the
performance predicted by our analytical model.
\end{abstract}

\section{Introduction}

Online Social Networks (OSN){\footnote{In this paper, we will also use another
term: Social Networking Services (SNS) to refer to not only OSN services, but
also other generic communication channels, e.g. RSS and Email, that can be
used to support online social interactions}} like Facebook and Twitter have
become an essential part of our daily life,
e.g. Facebook has over 1 billion\footnote{\url{http://newsroom.fb.com/download-media/4227}}
monthly active users -- 1/7 of the world population.
Despite
their overwhelming success, the centralized control of these services have led
to serious concerns about user privacy, censorship {\cite{wsp2013prism}} and
operational robustness. They are now obvious targets or even vehicles of many
totalitarian regimes which constantly seek to monitor and control information
dissemination among their people. Some even argues that the centralized and
absolute control power associated with OSN services have resulted in many
non-user-friendly management styles or policies, e.g. the real world cases
reported by IndieWebCamp{\footnote{\url{http://indiewebcamp.com/}}} participants. The
aforementioned concerns have motivated the active development of Decentralized
Social Networks (DSN) in the recent years with the goal to let users regain
better control over their personal data.

Although numerous DSN projects have been launched in the past few years, to
date, very few of them have managed to proceed beyond the prototyping or
paper-publishing stage. Even the most successful, often cited example of DSN,
namely, Diaspora {\cite{diaspora}}, has only about 0.4
million{\footnote{Sept 25, 2013, estimated by
\url{https://diasp.eu/stats.html}}} users, let alone active ones.
While
this may not be surprising given the wide range of challenges in implementing,
bootstrapping and operating a DSN {\cite{datta2010decentralized}}, we believe
that the biggest hurdle for the widespread adoption of DSN services is the
lack of a gradual transition path for a user to migrate to a DSN without
leaving behind a large portion of his/her friends who are likely to stick with
only existing OSNs for convenience. As such, even for the minority early
adopters who are determined to move to a DSN service for its enhanced privacy
protection and possibly richer functionalities, the ability to perform
cross-platform socialization will be critical. Towards this end, we propose
SNSAPI, a middleware which enables an end-user to aggregate and stitch
together all of his/her online social activities and become a node in a ``meta
social network'' {\cite{hu2013metasn}} as illustrated in Fig. \ref{fig:instantiated-metasn}.
In fact,
many existing SNS users already consciously or subconsciously ``stitch'' the
heterogeneous platforms together by selectively relaying messages across
different platforms after some manual filtering and editing. For example,
after reading a ``juicy'' gossip from a blog, you may forward it manually to
your personal friends on Facebook. Such manual cross-platform forwarding
operations can be viewed as the formation of a meta social network which
overlays on top of the existing SNS. Our objective is to provide the tools and
systems that can help users to better perform cross-platform socialization. We
expect those tools and systems can gradually detach users from existing
centralized OSNs and allows a smooth transition to the decentralized ones. In
summary, this paper has made the following technical contributions:

$\bullet$ After reviewing related work in Section \ref{sec:related-works}, we
analyze the challenges of building a DSN and propose a meta social networking approach in Section
\ref{sec:a-meta-social-networking-approach-for-decentralization}.

$\bullet$ As described in Section \ref{sec:current-design-and-implementation},
we have designed, implemented and released multiple iterations of an
open-source middleware called SNSAPI to support cross-platform socialization
over existing SNS. Sample applications built using SNSAPI are also presented
to demonstrate its flexibility and extensibility.

$\bullet$ Design choices, observations and maintenance experience of SNSAPI are
discussed in Section \ref{sec:design-maintenance-and-refactor-experience}.

$\bullet$ We have conducted an experimental deployment of a 6000-node DSN over
PlanetLab based on the SNSAPI middleware and measured its empirical
performance. As a comparison, we also provide an analytical model to
characterize the performance of the system with results presented in Section
\ref{sec:a-medium-scale-dsn-in-the-wild}.

$\bullet$ We conclude our findings and propose follow-up work for the future
in Section \ref{sec:conclusions}.

\section{Related Work}\label{sec:related-works}

To overcome the drawbacks of centralized SNS, Decentralized Social Networks
(DSN) such as Diaspora {\cite{diaspora}}, Musubi {\cite{dodson2012musubi}}
and OneSocialWeb {\cite{onesocialweb}} have recently been proposed and
implemented. According to our classification in Section
\ref{sec:a-meta-social-networking-approach-for-decentralization}, most works
are Distributed Social Network (DisSN). Two examples of federation protocols
are OneSocialWeb {\cite{onesocialweb}} and Ostatus
{\cite{w3c2013ostatus}}, which are not bound to any particular software
implementation. 
For the DisSNs, there are several different system
architectures. PrPl {\cite{seong2010prpl}} and Musubi
{\cite{dodson2012musubi}} are two examples of fully decentralized systems.
Each user corresponds to a DSN node and they only have the view of their
direct connections. Diaspora {\cite{diaspora}} is a super-node based
system. Users can setup their own ``pod'' (server) or register on other
``pods''. Peoplenet {\cite{motani2005peoplenet}} and Rflex
{\cite{liu2012spontaneous}} are systems combining heterogeneous communication
technologies. Peoplenet {\cite{motani2005peoplenet}} first routes messages to
the neighbourhood of the target using cellular infrastructures and then
leverages opportunistic forwarding to complete the last hop(s) of the
delivery. Rflex {\cite{liu2012spontaneous}} targets spontaneous group
messaging and transparently switches between the cloud backend and NFC/ D2D
connections. ePOST {\cite{mislove2006experiences}} and PeerSoN
{\cite{buchegger2009peerson}} are two DHT-based solutions. While ePOST
{\cite{mislove2006experiences}} follows DHT loop for message delivery, PeerSoN
{\cite{buchegger2009peerson}} only uses DHT to locate users before
establishing direct connections for data transfer. These systems and protocols
adopt a clean-slate design and have few or none provisions to support migration.
Although some systems have import functions from existing service providers, 
they still have the lock-in effect because
they do not give a common data structure abstraction to support convenient
export and inter-operation. 

In terms of the ability to perform cross-platform operation, there are many
existing aggregation services on the Internet. For example, IFTTT
{\cite{ifttt}} abstracts Internet-based information services as ``channels''
and allows users to define ``IF-This-happens-Then-do-That'' (IFTTT) forwarding
recipes. Yahoo Pipes {\cite{yahoo_pipes}} supports fewer platforms but allows
more sophisticated processing by designing a story-board of logical operations
(e.g. condition, loop, etc). Yoono {\cite{yoono}}, SocialOomph
{\cite{socialoomph}}, and Hootsuite {\cite{hootsuite}} are other example
services which provide partial interoperability between heterogeneous
platforms with some basic personalization functions. While these services are
useful for novice users, they have at least one of the following problems: 1)
Only configurable but not programmable, which severely limits their functions;
2) Non open-source and thus less reusable for automation;
3) Not readily extensible to other platforms.
The open-source projects, OpenSocial {\cite{open_social}}
provides an abstraction for different OSNs but it requires a steep learning
for simple tasks.
ThinkUp {\cite{thinkup}}, an open-source web service,
reads messages from several OSNs and stores them in a local
database to facilitate the mining of important information. 
However, it does
not provide a comprehensive abstraction of heterogeneous SNS like SNSAPI
and its
web service nature also requires more infrastructure support
(e.g. LAMP environment), making it hard to run under resource critical
environment, e.g. mobile devices.

\section{A Meta Social Networking Approach for
Decentralization}\label{sec:a-meta-social-networking-approach-for-decentralization}

\begin{figure}
    \centering
    \subfigure[DisSN]{
        {\includegraphics[width=0.3\linewidth]{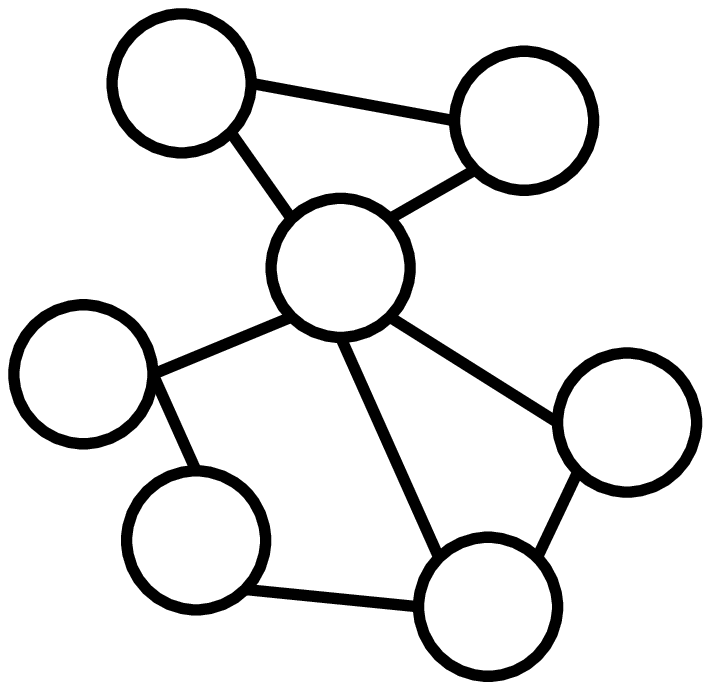}}
        \label{fig:dissn}
    }
    \subfigure[FedSN]{
        {\includegraphics[width=0.3\linewidth]{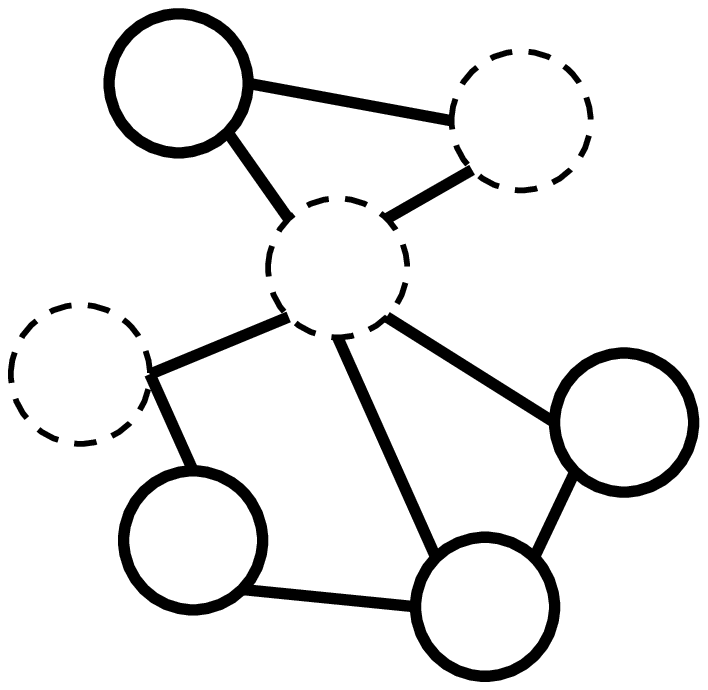}}
        \label{fig:fedsn}
    }
    \subfigure[MetaSN]{
        {\includegraphics[width=0.3\linewidth]{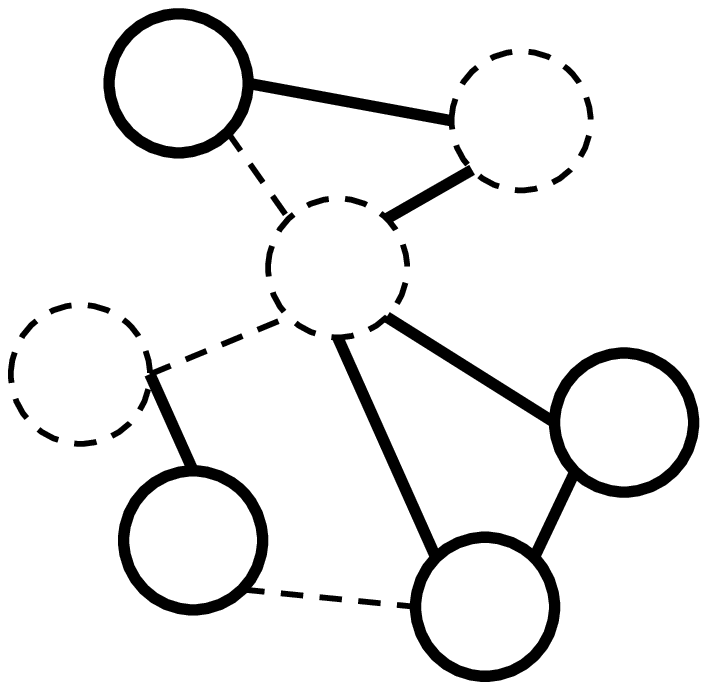}}
        \label{fig:metasn}
    }
    \caption{Three Different Approaches for Decentralized SNS:
        Distributed-, Federated-, Meta- Social Networks
    \label{fig:degree-of-decentralization}}
\end{figure}

Before delving into the details of our approach for the decentralization of
SNS, we briefly discuss three alternative paths towards this goal. According
to the degree of decentralization, there are three types of DSN as illustrated
in Fig. \ref{fig:degree-of-decentralization}:

$\bullet$ Distributed Social Network (DisSN). In a distributed social network,
nodes are homogeneous and runs the same software package -- the solid circles.
They use the same protocol -- the solid lines. Most DSN proposals and
implementations are actually DisSN, e.g. Diaspora {\cite{diaspora}},
Musubi {\cite{dodson2012musubi}} etc. In this sense, the DisSN approach is
{\tmstrong{software-package oriented}}.

$\bullet$ Federated Social Network (FedSN). Nodes in a FedSN can adopt
different software packages -- the ``solid'' and ``dashed'' circles in Fig. \ref{fig:fedsn}
as long as they support a common protocol, a.k.a. the federation protocol
represented by solid lines in Fig. \ref{fig:fedsn}.
Email and OStatus {\cite{w3c2013ostatus}} are two examples of this approach.
In short, FedSN is {\tmstrong{protocol oriented}}.

$\bullet$ Meta Social Network (MetaSN) {\cite{hu2013metasn}}. MetaSN is the
network of different social networks: In this case, 1) there are many
different software packages (server/ client implementations) and 2) nodes may
use different protocols -- the solid or dashed lines in Fig. \ref{fig:metasn}.
Information diffusion on MetaSN does
not rely on a single software package or a single protocol. Instead, it is
done via multi-hop bilateral communications. There is a common object to hold
information and it can have different representations on different links. In
other words, MetaSN is {\tmstrong{object oriented}}. One can see
that MetaSN is highly reminiscent of our real life social network in which
different people may talk different languages. Socialization is not dependent
on a single language but on multi-hop bilateral communications.

\begin{figure}
    \centering
    {\includegraphics[width=0.8\linewidth]{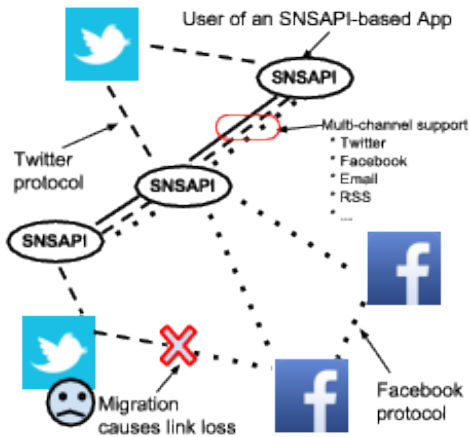}}
    \caption{The Incremental Deployment of Decentralized Social Network Based on SNSAPI}
    \label{fig:instantiated-metasn}
\end{figure}

We have designed a cross-platform middleware, SNSAPI, to realize the MetaSN
approach described above. Fig. \ref{fig:instantiated-metasn}
illustrates a MetaSN formed by SNSAPI and existing OSNs. Users running an
SNSAPI-based App can socialize with users of mainstream OSNs, e.g. Twitter
and Facebook, using the corresponding protocol. They can also conduct the same
types of online social activities with each other in a transparent manner via
other communication channels supported by SNSAPI, e.g. (E)mail and (R)SS. In
contrast, the Twitter user and the Facebook user at bottom of the figure
cannot communicate with each other due to the lack of a common protocol.
SNSAPI users also have the ability to aggregate, filter and/or relay messages
in this MetaSN. As such, the SNSAPI-based nodes can enable multi-hop social
interactions among users who are originally locked into their own silo of
existing OSNs. In short, the SNSAPI can help to jumpstart and then grow a DSN
in an incremental, gradual manner.

\section{SNSAPI Design and
Implementation}\label{sec:current-design-and-implementation}

Our cross-platform middleware, SNSAPI, is designed based on the following
principles: 1) Focusing on solving the 80\% problems; 2) Staying open to
support future service evolution ; 3) Trading execution performance for script
development efficiency.
In this section, we first introduce the overall architecture
and the core abstractions. After that, we present some sample applications to
show the flexibility and programmability of SNSAPI.

\subsection{Overall Architecture}

\begin{figure}[h]
    \centering
    {\includegraphics[width=0.9\linewidth]{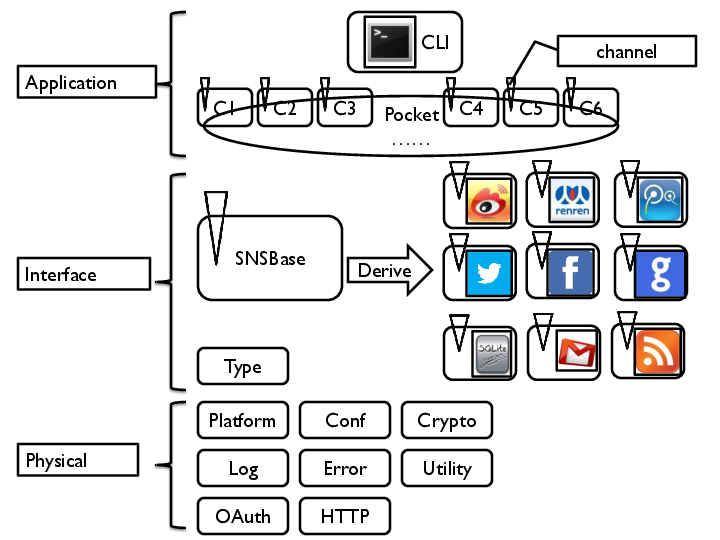}}
    \caption{The SNSAPI Architecture\label{fig:the-snsapi-architecture}}
\end{figure}

Figure \ref{fig:the-snsapi-architecture} depicts the architecture of SNSAPI,
which consists of the following three layers:

$\bullet$ Interface Layer (IL). \tmtexttt{SNSBase} is the base class for all
kinds of SNS and can be derived to implement real logic that interfaces with those platforms. 
In SNSAPI terminology, the module containing
derived classes is called ``plugin''; the derived class from
\tmtexttt{SNSBase} is called ``platform''; the instance of the class is called
``channel''. 
Message and message list classes are also defined in
this layer to provide the core abstraction to support MetaSN.

$\bullet$ Physical Layer (PL). There are many common operations when
interfacing with different SNS, e.g. HTTP, OAuth, error definition, etc.
We implement them in the PL so that plugins can reuse them. 
To enhance flexibility, we provide a wrapper class for
most 3rd party modules so that users can substitute (all or part of) them
with others without modifying the core of SNSAPI.

$\bullet$ Application Layer (AL). 
To reduce repeated works in batch operations, we developed a ``Pocket'' class in AL. 
It
is a container to hold multiple channels. 
For most applications, Pocket should
be the Service Access Point (SAP) to SNSAPI. 
In this way, end users can enable
new channels by simple configurations and no intervention from App developer is needed.

\subsection{Abstraction of Interfaces}\label{sec:abstraction-of-interfaces}

We abstract the interfaces of different SNS platforms to support a set of
common primitives. Since not all platforms have direct support for those
primitives, we often have to translate the functions in order to provide a
unified view to the upper layers. The three fundamental primitives and some
examples are as follows:

$\bullet$ \tmtexttt{auth} -- This name is short for either ``authorization''
or ``authentication''. 
For example, in order to access OSNs like Facebook, we
need to go through the OAuth flow. 
In order to access email platform, the
usual approach is to authenticate via username/ password. 

$\bullet$ \tmtexttt{home\_timleine} -- This function gets the latest messages
targeted to the user on a certain channel. For an OSN, it retrieves the home
timeline.
For Email platform, this function gets latest messages from
INBOX, which resembles the home timeline of OSNs. 
\tmtexttt{home\_timeline} is the basic ``read'' function for a platform.

$\bullet$ \tmtexttt{update} -- It provides the basic ``write'' function for a
platform. On an OSN, this can be a status update, blog update, or others,
depending on the platform. On RSS platform, this function writes a new entry to the feed. 

Although \tmtexttt{home\_timleine} and \tmtexttt{update} can be named as
``read'' and ``write'' from a modeling viewpoint we stick to the OSN
convention. 
This philosophy is adopted throughout SNSAPI: Instead of providing
a unification scheme from scratch and requiring other existing/ new OSN
platforms to follow (as in the case of OpenSocial {\cite{open_social}}), we
adapt SNSAPI to existing platforms and work out a common divisor. From the
basic primitives, two more ``write'' functions can be derived: reply and
forward. The definition is different across OSNs due to differences in the
positioning of each platform. The model of ``write'' operations observed on
existing OSNs are further discussed in Section
\ref{sec:the-model-of-write-in-osn-platforms}. Following are the definitions
within the context of SNSAPI:

$\bullet$ \tmtexttt{reply} -- By replying a message, the user is able to add
comments to the original message. If Bob replies a message posted by Alice,
she will get a notification in some way. Where to post the comment is not
regularized and can be different across platforms.

$\bullet$ \tmtexttt{forward} -- By forwarding a message, the user is also able
to add comments to the original message. If Bob forwards a message posted by
Alice, she may or may not get a notification, depending on the platform.
However, the forward message together with the original message will appear in
the message update list of Bob. Note that \tmtexttt{forward} can be readily
implemented via \tmtexttt{update} and we have already included this
cross-platform forwarding function in the base class. Plugins can implement
platform-specific forwarding functions to enrich the 
functionalities.

\subsection{Abstraction of Data
Structures}\label{sec:abstraction-of-data-structures}

As is discussed in Section
\ref{sec:a-meta-social-networking-approach-for-decentralization}, we need to
design a common object to allow the formation of MetaSN. 
Message is the most
important data structure SNS but comes in different forms, 
e.g. JSON object returned by the API of one OSN, 
RFC822 formatted texts
of an email, or an XML entry of a RSS feed. 
Even if we only consider
conventional OSNs, the JSON objects returned can still be quite different. 
In order to facilitate cross-platform operations, we abstract a
\tmtexttt{Message} object.
It has the following four components:

$\bullet$ \tmtexttt{MessageID} -- It contains sufficient information for one
to identify a \tmtexttt{Message} across platforms. 
It is only
designed to be used by SNSAPI plugins. Users are not supposed to tap into the
fields of \tmtexttt{MessageID}.

$\bullet$ Mandatory fields -- This includes \tmtexttt{userid},
\tmtexttt{username}, \tmtexttt{text}, \tmtexttt{time}, and
\tmtexttt{attachments}. Those fields are actually the ``Greatest Common
Divisor'' of different SNS platforms according to our year-long refactoring experience.
The \tmtexttt{attachments} field is used to convey
non-text-based information like URL, image, and video files and can be an
empty list.

$\bullet$ Optional fields -- Although mandatory fields provide a common base
to perform cross-platform socialization, we may sometimes expect a richer set
of functionalities. Examples are like the ``like count'' and ``share count'' on some OSNs. 
Those fields are also unified via re-formatting and re-assembling of the raw response from OSN providers. 
App developers should test the existence before using it since they are optional.

$\bullet$ Raw data -- This is the original data obtained from an OSN service
provider. It gives the largest amount of information. 
However, App developer
should refer to the manual of specific platforms in order to use it.

\subsection{Plugins}

Plugin implements the logic to enable transaction with any specific SNS
platform. A standard plugin implements the 5 interfaces defined in Section
\ref{sec:abstraction-of-interfaces}. It also performs essential data
conversions from raw response to the data structures regularized in Section
\ref{sec:abstraction-of-data-structures}. Following is the list of plugins we
have already developed:

$\bullet$ Renren, Sina Weibo, and Tencent Weibo -- We started with these
platforms, which are the three largest OSN services in mainland China. The
plugins use RESTful APIs of the corresponding OSN. 
Based on the experience of
developing these plugins, we have derived some common building blocks, e.g.
HTTP, OAuth, and put them in the Physical Layer (PL).

$\bullet$ RSS, Email and SQLite -- These three platforms are not regarded as
social networking services in the traditional sense. However, they are
fundamentally the same as the OSNs mentioned above, 
i.e. a means to read/ write messages,
so it is easy to make them SNSAPI-compliant.

$\bullet$ Twitter and Facebook -- These two platforms are developed based on
existing 3rd-party wrappers.
Their existence shows that one can easily adapt existing wrappers
to be SNSAPI-compliant, so as to enable seamless cross-platform operation on new platforms. 
Pros/ cons of using 3rd-party wrappers are further discussed in Section
\ref{sec:user-3rd-party-wrappers-for-plugins}.

\subsection{Applications}\label{sec:applications}

The so-called SNSCLI, a Command-Line Interface in form of a Python shell, has
been developed as an application (app) of SNSAPI. Some typical usages include:
1) use SNSCLI to interface with programmes/scripts written in other languages
via STDIN/ STDOUT; 2) support interactive debugging for a platform.
Researchers have also developed crawlers in SNSCLI using the Vertical
Interfaces (Section \ref{sec:vertical-interface-and-horizontal-interface}) to
facilitate the collection of data over large-scale OSNs.

\begin{figure}
    \centering
    \subfigure[SNSGUI]{
        {\includegraphics[width=0.37\linewidth]{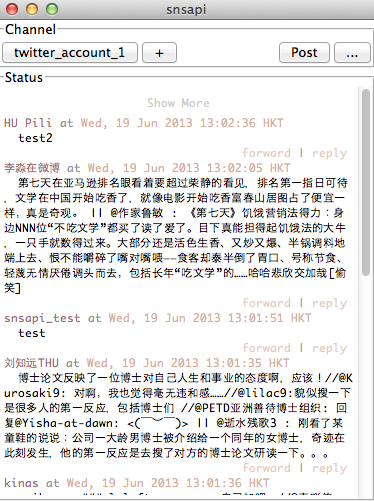}}
    }
    \hspace{1em}
    \subfigure[SNSDroid]{
        {\includegraphics[width=0.28\linewidth]{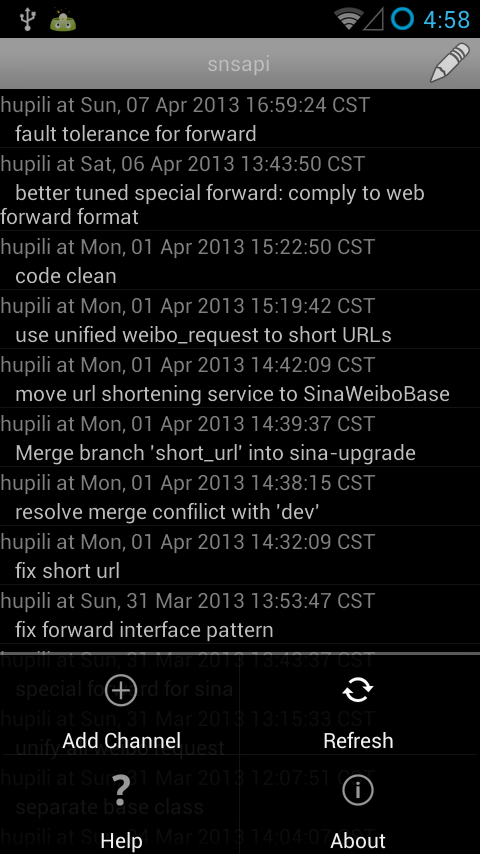}}
    }
    \caption{Screenshots of SNSGUI and SNSDroid
    \label{fig:screenshots-of-gui-on-pc-and-android}}
\end{figure}

After our third round of refactoring at the beginning of 2013, an open-source
contributor from GitHub, Tommy Alex, built a Graphical User Interface (GUI)
called SNSGUI (Fig. \ref{fig:screenshots-of-gui-on-pc-and-android} (a)) which
runs on major computing platforms including Linux, Windows, and OS X. Later
Tommy also ported SNSAPI to Android with a new UI (SNSDroid in Fig.
\ref{fig:screenshots-of-gui-on-pc-and-android} (b)). It is noteworthy that
Tommy managed to develop SNSGUI and SNSDroid in 10 days and 7 days,
respectively. Before developing SNSGUI, he has no prior knowledge of SNSAPI.
This demonstrates the flexibility and portability of SNSAPI.

We have also developed another App called SNSRouter {\cite{hu2013pixs}} which
provides a web UI and a ranking framework for cross-platform socialization.
Based on the unified abstraction of SNSAPI, the UI was developed in one week.
User efficiency is improved due to the convenient cross-platform operations
plus prioritization of incoming messages.
Among all kinds of benefits, it is worth to note that
SNSRouter can readily become a node in an ad hoc DSN by properly configuring
some channels supported by SNSAPI.

Due to space limit, we cannot present other existing Apps that have been
developed for SNSAPI, e.g. automatic forwarder, automatic replier, automatic
backup script. They can be found in our open-source repository. They all share
one distinguishing feature when comparing to their traditional counterparts:
these Apps all work in a platform-independent fashion and are ready to operate
on new platforms.

\section{Discussions on Design Choices and Key Observations}
\label{sec:design-maintenance-and-refactor-experience}

Since the initiation of SNSAPI project, we have gone through three major
rounds of refactoring. The project is still under active development and there
will be more forthcoming features. In this section, we provide a comprehensive
discussions of our experience in designing, maintaining and refactoring
SNSAPI. The philosophical shifts and key observations extracted from our
experience may shed light on future design of similar systems.

\subsection{Vertical and Horizontal
Interfaces}\label{sec:vertical-interface-and-horizontal-interface}

\begin{figure}
    \centering
    {\includegraphics[width=0.8\linewidth]{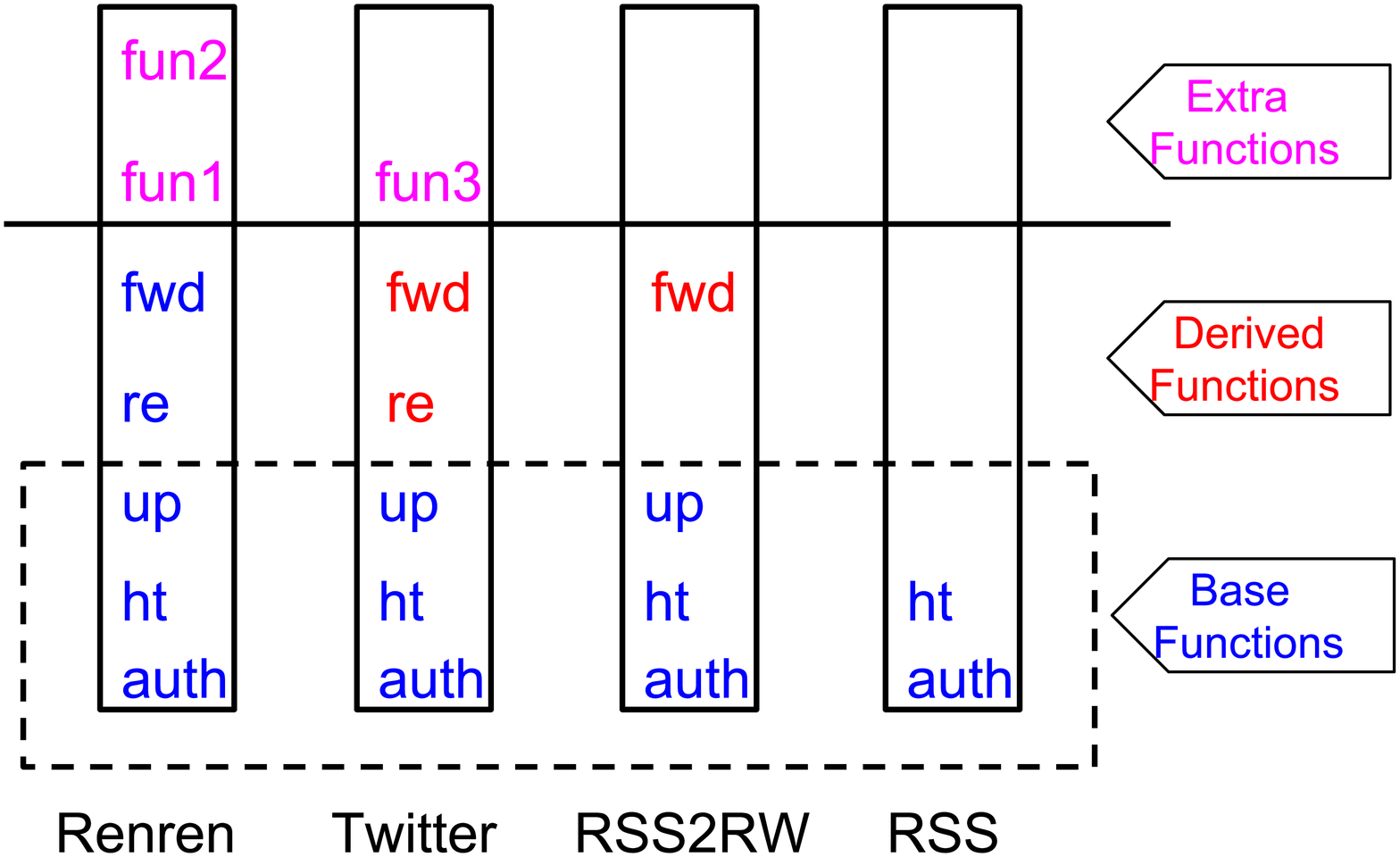}}
    \caption{Vertical and Horizontal Interfaces\label{fig:interface-horizontal-vertical.eps}}
\end{figure}

We categorize the functions provided by different social platforms into three
groups. As is described in Section \ref{sec:abstraction-of-interfaces}, base
functions include \tmtexttt{auth}, \tmtexttt{home\_timeline} and
\tmtexttt{update}; derived functions include \tmtexttt{reply} and
\tmtexttt{forward}. Any other OSN-specific functions are regarded as extra
functions, e.g. album list query and like. 
Extra functions are not very common across platforms, so we
do not try to unify them.
Fig.
\ref{fig:interface-horizontal-vertical.eps} illustrates the three types of
functions. Based on this categorization, we can deal the aforementioned
functions in a horizontal manner. 

While providing Horizontal Interfaces is the original goal of SNSAPI, we
found that some users expected SNSAPI to be a full-fledged wrapper for a given OSN platform.
Such mis-matched perception roots from the fact that there exists
many ad hoc wrappers in all kinds of languages for a wide range of OSN
platforms. 
We realized that the support of the Horizontal Interfaces alone will not
be adequate: in order to provide a unified set of horizontal interfaces, we
are forced to trim down existing operations in some platforms which results in
a loss of functionalities. To overcome this dilemma, we have to expose an
additional set of Vertical Interfaces which support a richer set of functions
but are allowed to be drastically different across various platforms. 

The Horizontal Interfaces can be used to systematically stitch
different social platforms together and they require a very mild learning
curve.
The Vertical Interfaces must be used with the assistance of
documents from service providers.
They are not the recommended way of using
SNSAPI because Apps have to deal with upcoming changes from the OSN providers.
Nevertheless, developers may find it useful when they are dealing with a specific platform,
e.g. large-scale crawling as is used in Section \ref{sec:a-medium-scale-dsn-in-the-wild}.

\subsection{The Model of ``Write'' on
OSNs}\label{sec:the-model-of-write-in-osn-platforms}

Although SNSAPI tried to conform to existing services from the very beginning,
the models provided by existing platforms may not be ultimately ideal.
We observed different models of ``write'' operations, i.e. {\tmsamp{update}},
\tmtexttt{reply} and \tmtexttt{forward}, from existing OSNs. A
summary of the write models is presented in Table
\ref{tab:write-models-of-different-osn-platforms} and we discuss the
advantages/ disadvantages as follows.

$\bullet$ Twitter: There is only one operation -- update. Twitter has a
``retweet'' built-in (Official ReTweet, ORT) but users find it
not very useful because they can not put comments while retweeting.
Towards this end, many Twitter users are still using the old user-invented
conventions (User-invented ReTweet, URT) by updating a status in the form:
\tmtexttt{RT @user message}.
This user invented ``RT'' notation is closer
to the ``forward'' operation defined in Section
\ref{sec:abstraction-of-interfaces}. If the ``\tmtexttt{@user}'' appears at
the beginning of a status update, it is regarded as a ``reply''. In other
words, the reply operation on Twitter is still an update operation in essence.

$\bullet$ Facebook: There are three operations regarding a status -- update,
share, and reply.
Facebook's reply is the same as our
definition in Section \ref{sec:abstraction-of-interfaces}. 
We will term this type
of reply as Facebook-Style Reply (FBSR), in contrast to the above mentioned
Twitter-Style Reply (TSR). 
FBSR is more privacy preserving 
but SNSAPI will lose track of those reply messages
because they do not enter the timeline of the replier.
When ``share'' one object on Facebook, users can add comments to it.
At first glance, it appears as a forward operation in our context. 
However, all share operations go to the original status and are threaded there. 
Compared with our notion of ``forward'', this ``share'' cannot track the forwarding traces.
In this case, it is better to textually construct a ``forward''' message by
the \tmtexttt{update} method directly (i.e. a URT) on Facebook. 
In fact, the general
\tmtexttt{forward} method in \tmtexttt{SNSBase} is designed to perform a URT
if the platform does not have a built-in ``forward'' function in our sense.

$\bullet$ Renren, Sina Weibo, Tencent Weibo: There are three ``write''
operations -- update, forward, and reply. 
In terms of product positioning,
Renren is a Facebook-like service in China; Sina Weibo and Tencent Weibo are
two Twitter-like ones. 
Interestingly, their write models are the same and
partly differ from both Twitter and Facebook as presented in Table
\ref{tab:write-models-of-different-osn-platforms}.

\begin{table}
  \centering
  \caption{Write models of different OSN
  platforms\label{tab:write-models-of-different-osn-platforms}}
  \vspace{1em}
  \begin{tabular}{|l|l|l|}
    \hline
     & reply & forward\\
    \hline
    Twitter  & update (TSR) & update (URT)\\
    \hline
    Facebook  & FBSR & N/A (``share'')\\
    \hline
    Renren  & FBSR & URT\\
    \hline
    Sina Weibo  & FBSR & URT\\
    \hline
    Tencent Weibo  & FBSR & URT\\
    \hline
    SNSAPI  & FBSR & URT\\
    \hline
  \end{tabular}
  \vspace{-2em}
\end{table}

\subsection{Pull Channel v.s. Push Channel}\label{sec:pull-channels-vs-push-channels}

We can abstract all SNS communications by directed graphs.
For undirected graphs, we just add both directions of a connection. 
This graph encodes the potential message flow,
i.e. the follower-followee relationships. 
One fundamental design
choice for a social network is whether to let a follower pull messages from a
followee or let a followee push messages to his/her followers. 
In the centralized scenario, this is only a matter of how database is synchronized.
The service exposed to users does not have pull/ push issues. 
In contrast, this design choice can result in very different resource usage under the
decentralized scenario.
In SNSAPI, both Pull-Channel models,
e.g. RSS, and Push-Channel models, e.g. Email, are supported. 
Since they have different properties,
it is desirable to choose the model on a per link basis,
rather than pin down a global setting.

\subsection{Use 3rd Party Wrappers for
Plugins}\label{sec:user-3rd-party-wrappers-for-plugins}

Before the launch of SNSAPI, there were already many ad hoc wrappers for
different OSN platforms.
Those wrappers are easy
to use as long as one only wants to deal with a specific platform. 
Whether to use 3rd party wrappers has been a design issue ever since the initiation of SNSAPI.
When we tried to add Twitter support to
SNSAPI, we found an already established 3rd party wrapper,
\tmtexttt{python-twitter}.
By adapting its interfaces to follow the SNSAPI convention, we
managed to support Twitter platform in half a day.
This shows that 3rd party wrappers can enable rapid prototyping for new
platforms and potentially make the upgrade process easier.
There are also some disadvantages like being less flexible for modification.
We have a qualitative
comparison shown in Table \ref{tab:using-3rd-party-wrappers}. 
As we can see, 
there is no definite answer for this design issue at present. SNSAPI,
together with the community, is still actively evolving and trying to test out
those alternative solutions.

\begin{table}
    \centering
  \caption{Pros and Cons of Using 3rd Party Wrappers 
  \label{tab:using-3rd-party-wrappers}}
      \vspace{1em}
  \begin{tabular}{|l|l|l|}
    \hline
    & Use & Not Use\\
    \hline
    Prototype efficiency & ***** & **\\
    \hline
    Upgrade efficiency & **** & ***\\
    \hline
    Interface stability & **** & **\\
    \hline
    Data structure stability & ** & **\\
    \hline
    Project Cleanness & * & ****\\
    \hline
    Degree of Freedom & ** & ****\\
    \hline
    Execution efficiency & **** & *****\\
    \hline
    Reliability & ** & ****\\
    \hline
  \end{tabular}
  \\
  (the more *'s the better)
      \vspace{-2em}
\end{table}

\subsection{Friend-List Management}\label{sec:friend-list-management}

Friend-list management for OSNs is so far only available via platform-specific Vertical
Interfaces (Section \ref{sec:vertical-interface-and-horizontal-interface}).
As the designers of SNSAPI, we have two strong reasons not to horizontally abstract
friend-list management functions.
Firstly, 
friend-list management is not a common and frequent operation
(so-called 80\% problems), 
so Out-Of-Band (OOB) management is acceptable.
Secondly, the Channel-list in SNSAPI is actually an implementation of
friend-list under the fully decentralized scenario.
Consider a social network formed using the RSS platform (Section
\ref{sec:a-medium-scale-dsn-in-the-wild}). 
One user can update status to a RSS
feed and publish it to some common location known by his/ her friends.
The other users can instantiate a RSS channel with the URL pointing to this location.
In this way, the SNSAPI users form a DSN and the channel-list of SNSAPI is actually the friend-list.
Instead of unifying friend-list management functions on different OSNs,
a better way is to build a channel management application on SNSAPI
to provide a more elegant abstraction.

\subsection{Use Existing Services as Aggregators}
\label{sec:use-of-existing-services-as-aggregators}

Although our primary goal is to evolve towards a decentralized social networks
structure, centralized OSN platforms are (and will remain to be) very
important components under the bigger Meta Social Network paradigm. As
discussed before, one reason is that those centralized OSN services are
already established and we must avoid link loss during the migration process.
Another reason is that centralized OSN platforms can actually serve as good
message aggregators.
Consider a special platform we built in SNSAPI --
\tmtexttt{RenrenStatusDirect}. With this platform, one can directly configure
the list of users he/she wants to follow.
Without explicitly adding the target
parties as friends via the request/ approval process, the SNSAPI user can
readily get their public status updates.
This approach gives us a clean model for friend-list management as is discussed in 
Section \ref{sec:friend-list-management}.
However, more resource is used for querying those ``friends'' one by one. 
The alternative approach, adding those friends
on an OSN first and configure only one normal channel in SNSAPI,
is more resource saving because a single query is enough to retrieve an aggregated timeline.

\section{Deployment of a 6000-node DSN on
PlanetLab}\label{sec:a-medium-scale-dsn-in-the-wild}

This section describes our experimental deployment of a large-scale DSN on PlanetLab
using SNSAPI. After instantiating a network of 6000 SNSAPI-based nodes on PlanetLab, 
we replay real-world update/ forward activity traces collected from Sina Weibo, 
one of the top OSNs in China\footnote{Incidentally, the
crawling of Sina Weibo for trace data is performed by another App of SNSAPI.}.
To our best knowledge, this is not only the largest experimental validation and systematic evaluation of any
DSN solutions in the literature but also the first one which involves the replaying 
of real OSN traces in this scale.

\subsection{Architecture of the Experiment}

Our experiment platform has three major components:

$\bullet$ SNSAPI: The cross-platform middleware described in Section
\ref{sec:current-design-and-implementation} 
-- the basic building block for this DSN.

$\bullet$ Bot: When the experiment starts, a bot running on each instance of a
SNSAPI-based node will continuously perform
four actions: 1) Periodically query timeline from their friends
(\tmtexttt{home\_timeline}); 2) Update SNS status according to a schedule
(\tmtexttt{update}); 3) Forward status according to a data-driven user behavior model
to be described in Section \ref{sec:data-driven-user-action-model}
(\tmtexttt{forward}); 4) Monitor nodal resource usage. 
We define an important system
parameter called Query Gap (QG) which is the interval between two rounds of the timeline query action.
 In our experiment, we use a system-wide constant value for QG and 
 show that this simple strategy suffices for the initial
formation of a medium sized DSN (e.g. 6000 nodes). To avoid synchronization
problems, every bot sleeps for a uniformly random duration between 0 and QG before the
issuing its first timeline-query to its neighbours.

$\bullet$ PlanetLab Manager (PLM): PlanetLab is a global research infrastructure
widely used to evaluate new networking systems. PlanetLab nodes provide no guarantee
on their service availability and can go online/offline at any time. 
This is ideal to evaluate the robustness and
elasticity of our proposed DSN architecture. 
We have also developed a toolbox to manage our
experiment on PL which includes modules for 1) base environment
deployment such as the nodal installation of Python2.7 and
virtualenv{\footnote{http://www.virtualenv.org/}}; 2) Node-screening which
help us to avoid machines with large time-drift and those which are behind
incompatible firewall ; 
3) Experiment workflow management
like workspace cleaning, bot distribution, initialization/ starting/ terminating each experiment, 
and nodal log collection.

In principle, we can connect our bots through all kinds of platforms supported
by SNSAPI. In reality, most OSN platforms require user's intervention before
SNSAPI can access it: App-key application, configuration on the OSN, and
completion of an authorization flow. To allow automatic deployment of
thousands of bots, we choose to use RSS in this experiment.
For practical deployment in the wild, each
pair of SNSAPI users can choose their preferred communication channel as our MetaSN model
only requires bilateral communications.

\subsection{Trace Collection from a Sina Weibo}

We collected real data from Sina Weibo, the largest microblogging service in
China, to evaluate the proposed ad hoc DSN based on SNSAPI. 
We leverage the
Vertical Interface (Section
\ref{sec:vertical-interface-and-horizontal-interface}) of SNSAPI to crawl the
profile, status update, forwarding history, as well as the followee list of each user. 
In order to get a meaningful trace to evaluate our system, we need a tightly knitted
group of active users with content frequently originated from and forwarded
within the group. The following heuristic to used to identify such group of candidate
users: 1) We use 60+ users categorized as ``programmers'' by Sina Weibo as
our seed{\footnote{http://verified.weibo.com/fame/itchengxuyuan/?srt=4}}; 2)
From the forwarding history of these seed users, we extract 6000+ more users.
The overall development plus execution time of this SNSAPI-based Sina Weibo
crawler is less than two days. This shows the remarkable development and
execution efficiency provided by SNSAPI. In total, we collected 33GB of raw JSON
data, which includes 12 million status update/forward actions, involving 
480,000 users in the forwarding chain.

\subsection{Data-Driven User Behavior
Model}\label{sec:data-driven-user-action-model}

One of the primary reasons for users to use OSN is to identify and disseminate
interesting messages. Using only passive measurements, it is impossible to
determine what messages are seen and what messages are interesting to each
user. As a surrogate, we consider the forwarding behaviors, which are strong
indicators of user interest and message relevance. A key objective is to determine 
the additional forwarding latency caused by our SNSAPI-based DSN
when comparing to that of a centralized OSN. 

\begin{figure}[h]
    \centering
    {\includegraphics[width=0.9\linewidth]{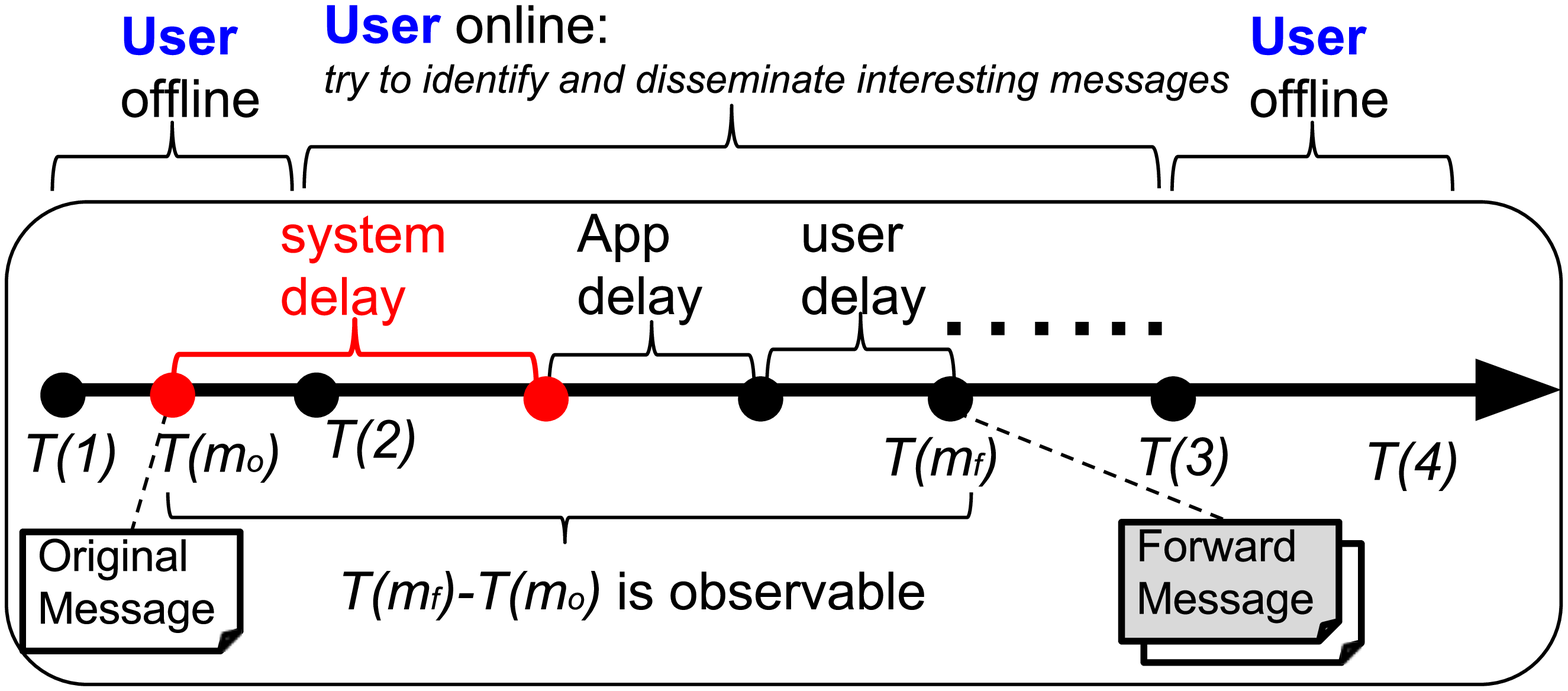}}
    \caption{Illustration of Real User Action Cycle\label{fig:user-model.eps}}
\end{figure}

Consider the process illustrated in Fig. \ref{fig:user-model.eps}: User
$u_o$ post the original message $m_o$ at $T (m_o)$; User $u_f$ forward this
message by posting $m_f$ at $T (m_f)$. The following factors will affect the end-to-end
delay from $T (m_o)$ to $T (m_f)$:

$\bullet$ User online/offline pattern: From $T (1)$ to $T (2)$ the user is
offline. The original message $m_o$ is posted during this period but cannot
be forwarded by $u_f$.

$\bullet$ System delay: This is the time required by $m_o$ to go from $u_o$'s
database to $u_f$'s database. For a centralized OSN, there is conceptually
only one database so this delay component should be negligible. 
That is, as soon as $u_o$ post $m_o$, $u_f$ is able to get $m_o$ in his/her database. 
For DSN, it can take considerable time for $m_o$ to go to $u_f$'s database, which may cause
a major increase in latency.

$\bullet$ Application delay: Users rely on certain application to retrieve
their home timeline, e.g. the web UI or the mobile App of Facebook. Additional
delay is added due to various reasons like refreshing frequency or message
re-ranking strategy. Users rely on many different applications to access
social networks, so this part of delay can be highly variant.

$\bullet$ User delay: After the above stages, $u_f$ already has $m_o$ in the
home timeline. In order to identify the interesting messages, $u_f$ will spend
time on reading the messages, watching attached videos, or even following the
embedded links. Additional delay is added until $m_f$ is finally posted at
time $T (m_f)$.

Our analysis of the user action cycle shows that: 1) System
delay causes the major difference between centralized and decentralized social
networks; 2) Factors like application delay and user delay are too complex to
be characterised by simple models; 3) Precise user online/offline pattern is only
available to the SNS providers; 4) The end-to-end delay, i.e. $T (m_f) - T(m_o)$, 
is observable via passive measurement. Towards this end, we adopt an
overall data-driven model, established based on two assumptions: 1) A bot
is always online, trying to identify and disseminate interesting messages; 2)
There is an intrinsic delay for each message to be forwarded. We use $T (m_f)
- T (m_o)$ to model the intrinsic delay for each message. Suppose one bot in
our DSN see the message $m_o$ at time $t$, then the forward time $T (m_{f'})$
is determined by the two rules:

$\bullet$ 1) If $t \leqslant T (m_f)$, let $T (m_{f'}) = T (m_f)$

$\bullet$ 2) If $t > T (m_f)$, let $T (m_{f'}) = t$

\begin{figure}[h]
    \centering
    {\includegraphics[width=0.9\linewidth]{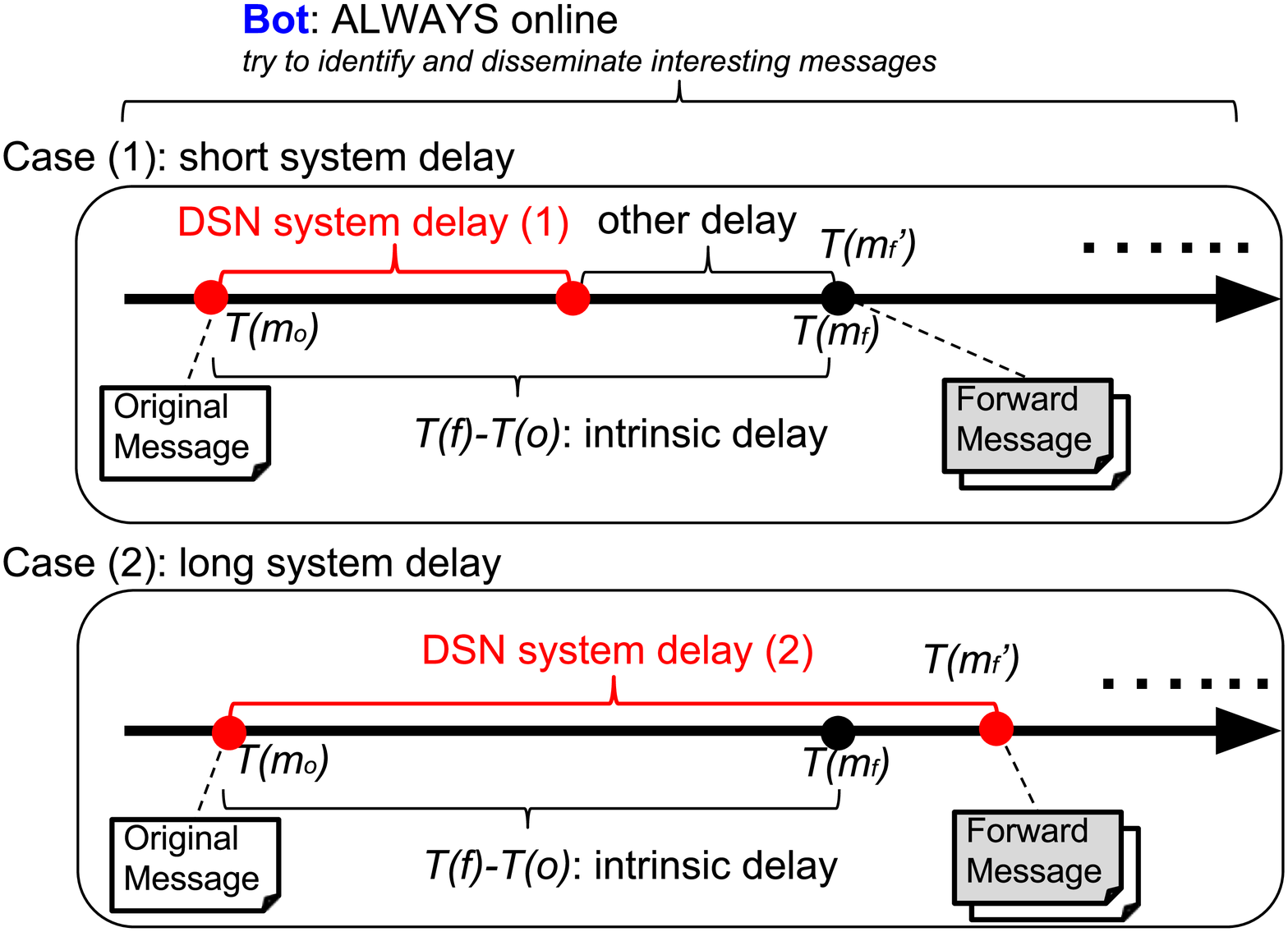}}
    \caption{Data-driven User Model: Bot + Intrinsic
    Delay\label{fig:user-model-bot.eps}}
\end{figure}

The first rule is to respect the delay observed from real traces. If our DSN
delivers messages quickly enough, then the bot waits until the intrinsic delay
is passed. The second rule comes from the bot assumption. Since the bot is
always online and the message is seen after the intrinsic delay period, it
forwards the message immediately. Fig. \ref{fig:user-model-bot.eps} 
depicts this data-driven user behavior model. Based on this model, 
our core performance metric is the Extra Forwarding Delay (EFD),
defined as $T (m_{f'}) - T (m_f)$.

\subsection{Experiment Setup}\label{sec:a-case-study-of-6k-node-dsn-based-on-snsapi}

To launch our experiment, we first distribute 6733 bots to 450
PlanetLab nodes randomly. The experiment lasts for one day wall-clock time. We
filter out update and forward activities of those users from Aug 1, 2013
0:00:00 to Aug 1, 2013 23:59:59 (UTC). During this period, there are 1168
messages originated by the 6733 Sina Weibo users and are further forwarded within
this group. The Query Gap (QG) is empirically set to 5 minutes. In real life,
users can manually trigger the \tmtexttt{home\_timeline}
function if one wants to get the updates of his friends immediately.

\begin{figure}
    \centering
    \subfigure[CPU]{
        {\includegraphics[width=0.46\linewidth]{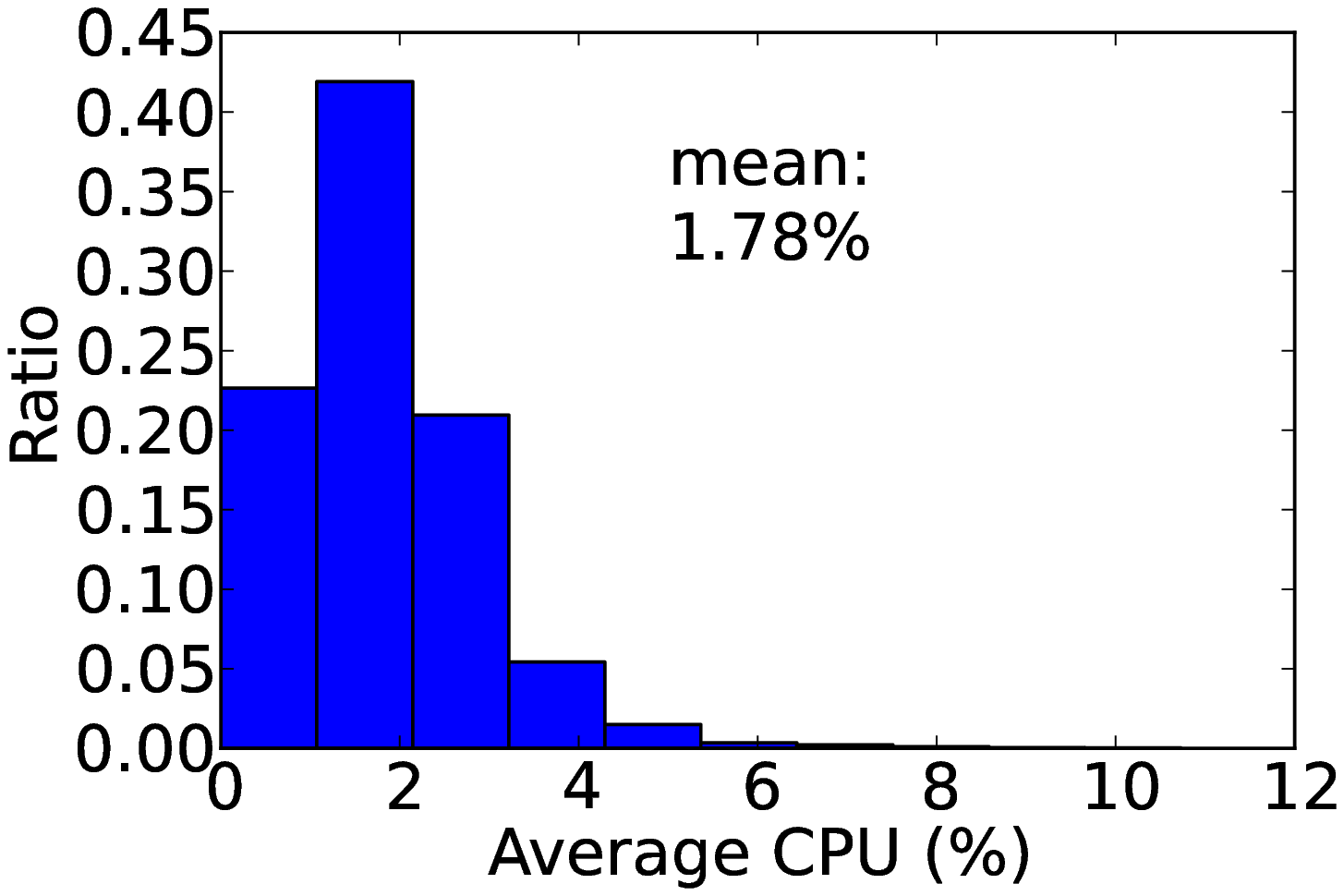}}
        {\label{fig:avg_cpu_hist.eps}}
    }
    \subfigure[Memory]{
        {\includegraphics[width=0.46\linewidth]{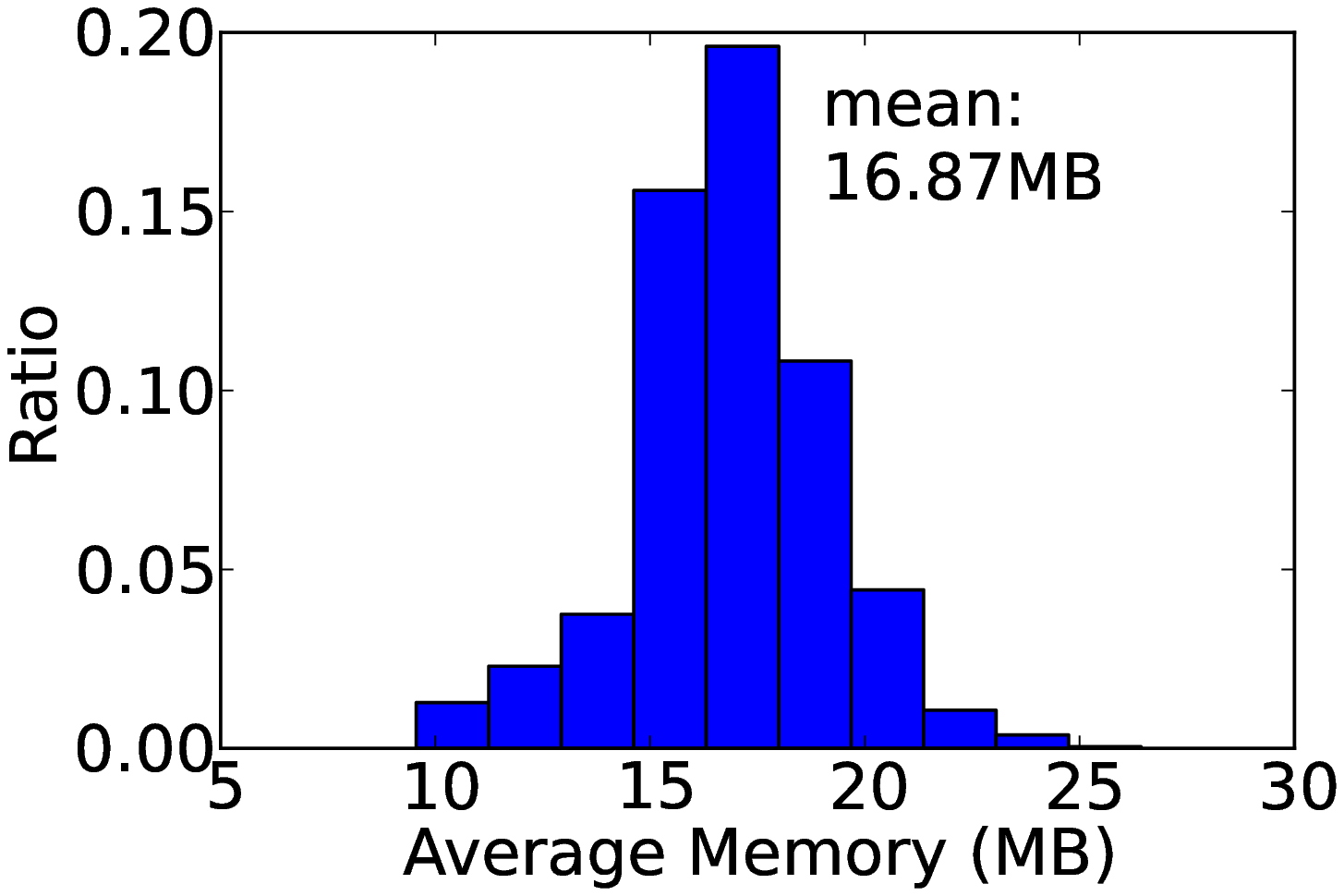}}
        {\label{fig:avg_mem_hist.eps}}
    }
    \subfigure[HTTP Query]{
    {\includegraphics[width=0.46\linewidth]{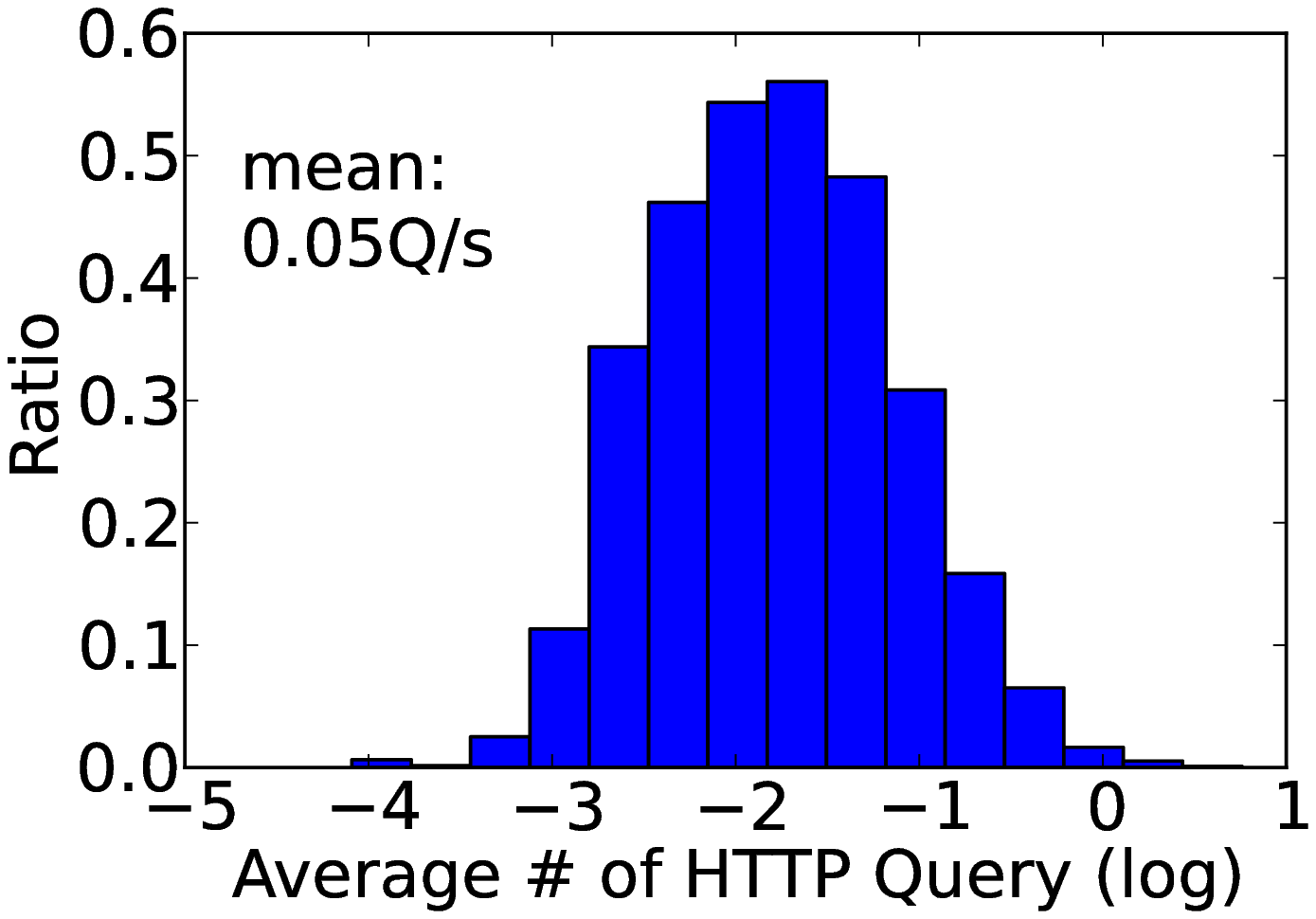}}
    {\label{fig:avg_http_query_hist.eps}}
    }
    \subfigure[HTTP Size]{
        {\includegraphics[width=0.46\linewidth]{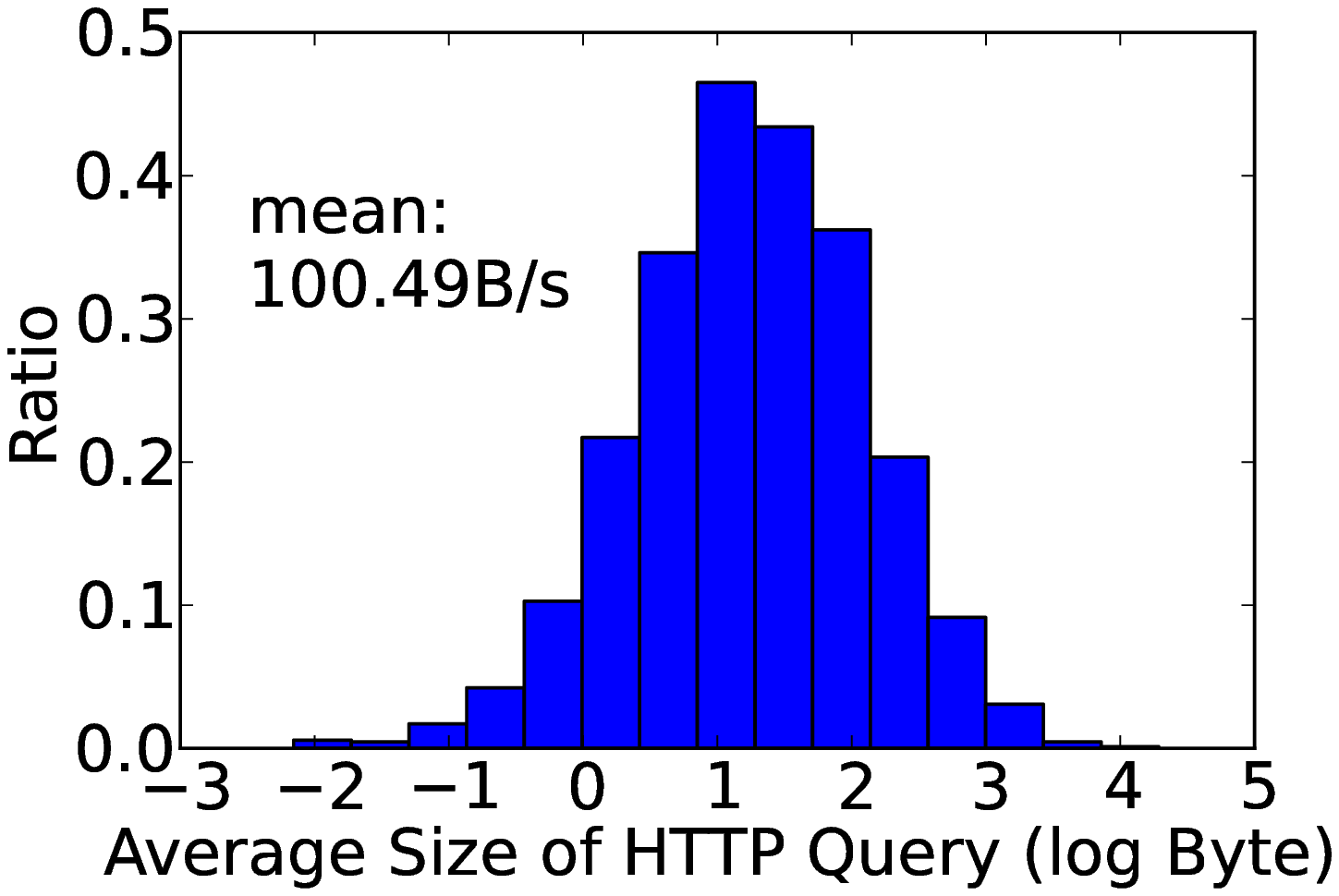}}
        {\label{fig:avg_http_size_hist.eps}}
    }
    \subfigure[RSS]{
        {\includegraphics[width=0.46\linewidth]{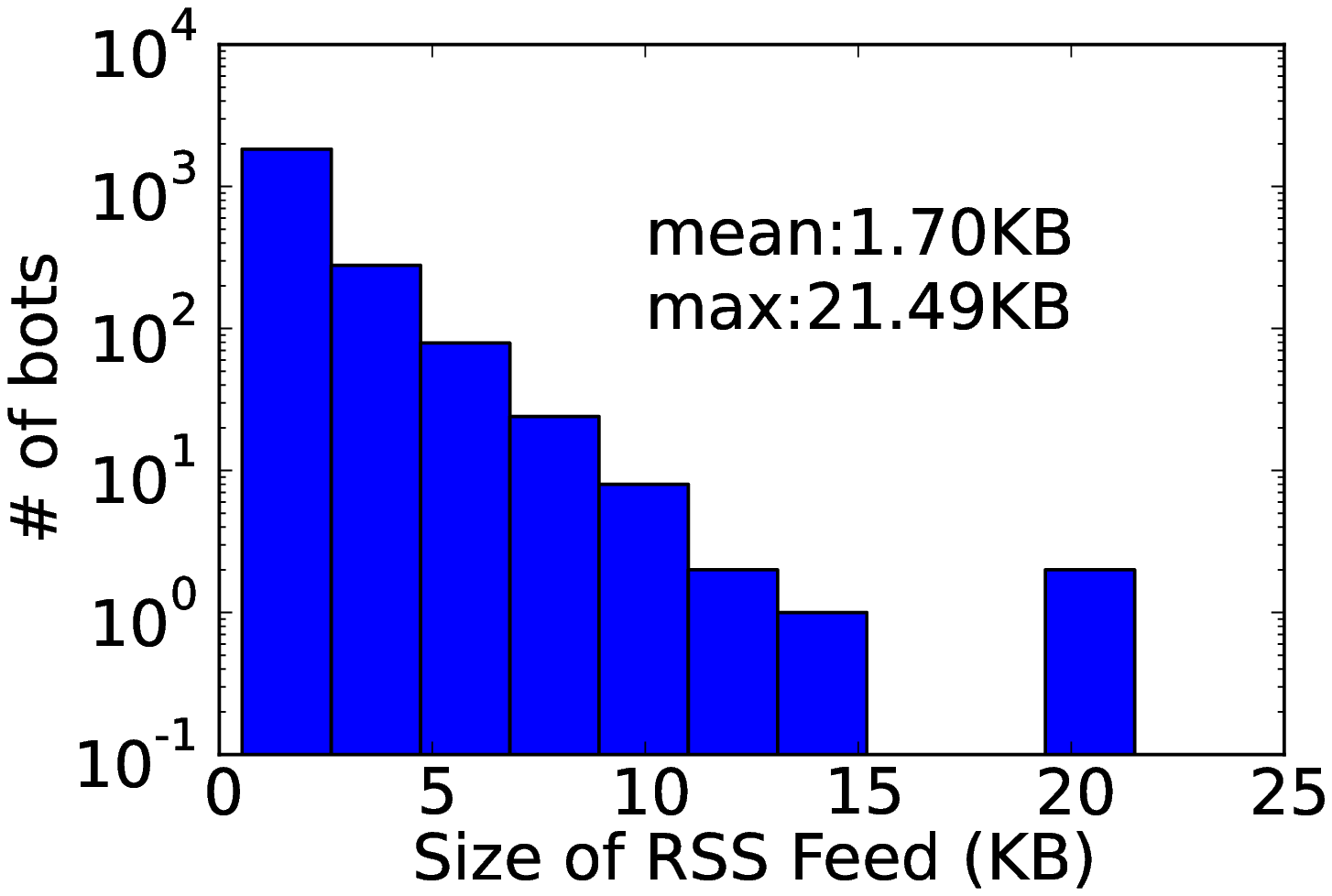}}
        {\label{fig:atom_size_hist.eps}}
    }
    \subfigure[SQLite]{
        {\includegraphics[width=0.46\linewidth]{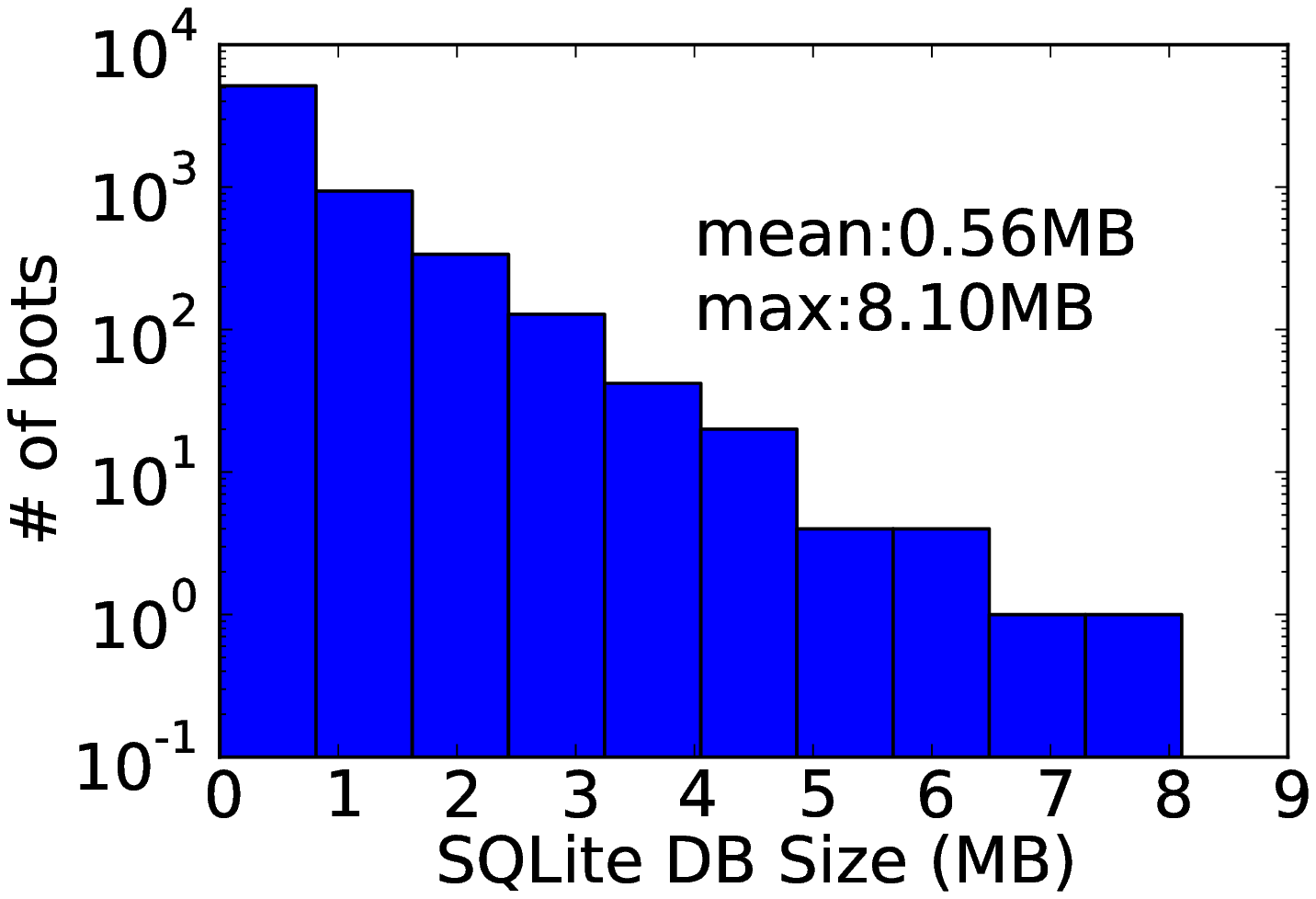}}
        {\label{fig:sqlite_db_size_hist.eps}}
    }
    \caption{Consumption of Different Resources}
\end{figure}

After a replaying a 24-hour real trace, we collect 57GB of logs from the PlanetLab
nodes. We then compute the average resources consumed by each bot. The histograms
and summary statistics are presented from Fig. \ref{fig:avg_cpu_hist.eps} to
Fig. \ref{fig:sqlite_db_size_hist.eps}. Notice that the nodal memory
usage includes both the bot and SNSAPI. Without the extra bot logic, the typical
value is 10MB, i.e. for running the SNSCLI. There are two major disk-space consumption
in this experiment. First is the RSS feed file: Each bot writes its feed to a
directory which is then served by a lightweight HTTP server. Second is a SQLite
database to support asynchronous operations via the buffering of incoming and
outgoing messages. From Fig. \ref{fig:atom_size_hist.eps} and Fig.
\ref{fig:sqlite_db_size_hist.eps}, we see an exponential decay and the typical
values are smaller enough to hold in the RAM so that disk storage and access
can be avoided in real applications.

\begin{figure}[h]
    \centering
    {\includegraphics[width=0.6\linewidth]{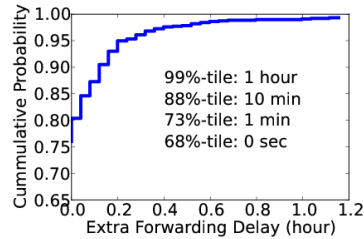}}
    \caption{c.d.f of Extra Forwarding Delay\label{efd_cdf.eps}}
\end{figure}

The c.d.f. and some percentiles of Extra Forwarding Delay (EFD) are shown Fig.
\ref{efd_cdf.eps}. We see that 68\%, 73\%, 88\% and 99\% messages are
forwarded with no EFD, $<$ 1min EFD, $<$ 10min EFD and $<$ 1hour EFD, respectively.
Since social networking is not a real-time service and there are already
considerable delay other than system delay, we remark that this EFD is
acceptable.

The mild resource consumption and small EFD shows that SNSAPI is a viable
solution to enable rapid and incremental deployment of an ad hoc DSN.
Furthermore, users can run SNSAPI not only on all major desktop operating systems (as demonstrated in Section \ref{sec:applications})
but also resource-constrained devices like smartphones.

\subsection{Analytical Model for the Pull-based
Backbone}\label{sec:analytical-model-for-the-pull-based-backbone}

Although SNSAPI supports both pull and push channels (Section
\ref{sec:pull-channels-vs-push-channels}), we only used pull channels, namely
RSS, in our experiment. The major parameter for this pull-based DSN is the
timeline Query Gap (QG) and we denote it by $h$. We want to characterise the 
relation between $h$ and the following random variables (r.v.):

$\bullet$ $R$: An abstract notation for certain resource. It is an r.v.
parameterized by $h$. The expectation of $R$ is inversely proportional to $h$.
Without loss of generality, we let $\Epsilon [R] = \frac{1}{h}$.

$\bullet$ $C$: CPU usage. $C = C_b + \beta_C R$. $C_b$ is the CPU usage of the bot
logic, i.e. checking and forwarding messages every second. It is constant w.r.t. the
poll parameter $h$. $\beta_C R$ is the resource consumption of SNSAPI.
$\beta_C$ is the coefficient to be fit later. Taking expectation, we have the
relationship: $\Epsilon [C] = \Epsilon [C_b] + \beta_C \Epsilon [R] = \Epsilon
[C_b] + \frac{1}{h} \beta_C$. However, we already know from the system design
that bot logic will take up the major portion of CPU usage. We expect
$\beta_C$ to be a near-zero coefficient.

$\bullet$ $M$: Memory usage. As $h$ varies, only query frequency is changed.
Memory consumption is not affected by it. It is a r.v. independent of $R$ and
$h$.

$\bullet$ $D$: DISK I/O. $D = \beta_D R$. There are two parts of major disk
I/O as discussed in Section
\ref{sec:a-case-study-of-6k-node-dsn-based-on-snsapi}. The maximum size of RSS
feed and SQLite DB are 20KB and 8MB, respectively. In real applications, they
can be fully held in RAM, thus eliminating all disk I/O. Since this only adds a
constant to $M$, we do not build an individual model for disk I/O.

$\bullet$ $H$: HTTP usage. In our experiment, the major network I/O is caused
by HTTP. Others like domain name lookup only create very little network load. We
measure HTTP usage from the followee's point of view. $H_q$ is the number of
HTTP query one bot receives and $H_s$ is the size of HTTP query it serves.
Since the traces are from a micro-blogging service, the size of messages are
upper bounded. Thus, we can assume $H_s \propto H_q$ and formulate the
following relationships: $\Epsilon [H_s] = \beta_{H_s} \Epsilon [R] =
\frac{1}{h} \beta_{H_s}$ and $\Epsilon [H_q] = \beta_{H_q} \Epsilon [R] =
\frac{1}{h} \beta_{H_q}$.

$\bullet$ $\Delta$: The delay from the original message to the forward
message in our DSN system. This depends on $I$, the intrinsic delay.

Most of the relationships are clear and ready for model fitting. The
end-to-end forwarding delay $\Delta$ can be characterized based on
the following additional assumptions:

$\bullet$ A1: $h$ is the dominant delay component in this system so that other
small delays can be neglected.

$\bullet$ A2: We assume an ideal Internet connection. Given current typical
access rate and the size of RSS feeds in Fig. \ref{fig:atom_size_hist.eps},
one bot can finish the query to all its neighbours within seconds. This is
also very small compared to typical values of $h$ so that we can further assume
one round of query can be completed immediately upon its issuing.

In reality, the forwarding traces form a tree. Since our target is $\Epsilon
[\Delta]$, we can invoke the linearity of expectation, namely $\Epsilon
[\Delta]$ is the summation of per-segment delay. In this way, we essentially
decompose the tree into individual forwarding chains. Denote the length of
forwarding sequence by $L$, which will be fit later using data from real traces. 
Then the expectation of end-to-end delay can be expressed as: (total expectation)
\[ \Epsilon [\Delta] = \int \Epsilon [\Delta |L = l] \Pr \{ L = l \} \mathd l
\]

Next, we focus on the analysis of a chain of length $l$. We denote the
original message as $m_0$ and the forward messages as $m_1, \ldots, m_l$. We
denote the absolute time when message $m_k$ is posted as $T (m_k)$. Let
$\Delta_k$ be the per-segment delay so that $\Delta_k = T (m_k) - T (m_{k - 1})$
according to definitions. The per-segment intrinsic delay is denoted as $I_k$.

Denote the absolute time that a bot starts to pull messages by $H$. Note that
the bots sleep a uniformly random duration before the first pull. Then we have $(t
- H) \tmop{mod} h \sim U [0, h]$ for any fixed time $t \in \mathbbm{R}$. We
use the deferred decision principle to study the forwarding behaviour of
$m_k$. That is, we first fix $T (m_{k - 1}) = T (m_0) + \sum_{j = 1}^{k - 1}
\Delta_k$, and defer the realization of r.v. $H$. $(T (m_{k - 1}) - H)$ is the
duration from first pull to the post of $m_{k - 1}$. By modular $h$, we get
the time elapse from last pull to the post of $m_{k - 1}$. Then $W = h - ((T
(m_{k - 1}) - H) \tmop{mod} h)$ is the waiting time until next pull and we
have a compact expression for $\Delta_k$ according to the data-driven user
behavior model described in Section \ref{sec:data-driven-user-action-model}:
\[ \Delta_k = \max \{ I_k, h - ((T (m_{k - 1}) - H) \tmop{mod} h) \} \]
If we fix $T (m_{k - 1})$, $(T (m_{k - 1}) - H) \tmop{mod} h \sim U [0, h]$
and further $W = h - ((T (m_{k - 1}) - H) \tmop{mod} h) \sim U [0, h]$. Then
the overall delay conditioned on the length being $l$ is:
\[ \Delta \middle|_{L = l} = \sum_{k = 1}^l \Delta_k = \sum_{k = 1}^l \max \{
   I_k, W \} \]
We now calculate the Extra Forwarding Delay (EFD), denoted by
$\Delta_{\tmop{EFD}}$. This is to use end-to-end delay in our system minus
intrinsic delay (the delay expected in a centralized system):
\[ (\Delta_{\tmop{EFD}}) |_{L = l} = (\Delta - I) |_{L = l} = \sum_{k = 1}^l
   \Delta_k - \sum_{k = 1}^l I_k = \sum_{k = 1}^l [\Delta_k - I_k] \]
Taking expectation, we have:
\[ \Epsilon [\Delta_{\tmop{EFD}} |L = l] = \sum_{k = 1}^l \Epsilon [\Delta_k -
   I_k] = \sum_{k = 1}^l \Epsilon [\max \{ I_k, W \} - I_k] \]
where $W$ is a uniform random variable in $[0, h]$. The overall EFD is:
\[ \Epsilon [\Delta_{\tmop{EFD}}] = \int \Epsilon [\Delta_{\tmop{EFD}} |L = l]
   \Pr \{ L = l \} \mathd l \]
We assume $I_k$ to be i.i.d. and use $I$ for a shorthand. Then the expression
is simplified to:
\[ \Epsilon [\Delta_{\tmop{EFD}}] = \Epsilon [L] \Epsilon [\max \{ I_k, W \} -
   I_k] \]
We can fit $I$ and $L$ from real traces and give the final expression for EFD.

\subsection{Model Evaluation using Experimental results}

In this section, we run 20 experiments parameterized by $h$ and compare
 the corresponding empirical measurement results with those predicted by
the analytical model. We replay 5 days worth of traces
from Aug 1, 2013 0:00:00 to Aug 5, 2013 23:59:59 (UTC) to drive the experiment
in order to reduce diurnal variation. 
To make the experiment tractable and also reduce PlanetLab machine variance, we accelerate
the experiment time by 12 times. That is, the 5-day traces are replayed in 10
real-life hours. We also trim the DSN topology to 4881 nodes by excluding those nodes
having no update or forward activities during this period.\footnote{In previous
experiments, each PlanetLab machine hosts about 15 bots and our slice was
killed on some machines due to the high aggregated memory usage (200-300MB). Since
PlanetLab is a shared environment, the killing decision also depends on
how other users are sharing the machine. After reducing of the DSN to 4881 nodes, 
we managed to stay in the ``safe region'' of memory consumption during the two-week long experiment period.}

\subsubsection{Fitting Intrinsic Delay and Forwarding Sequence Length}

\begin{figure}
    \subfigure[Intrinsic Delay]{
        \includegraphics[width=0.46\linewidth]{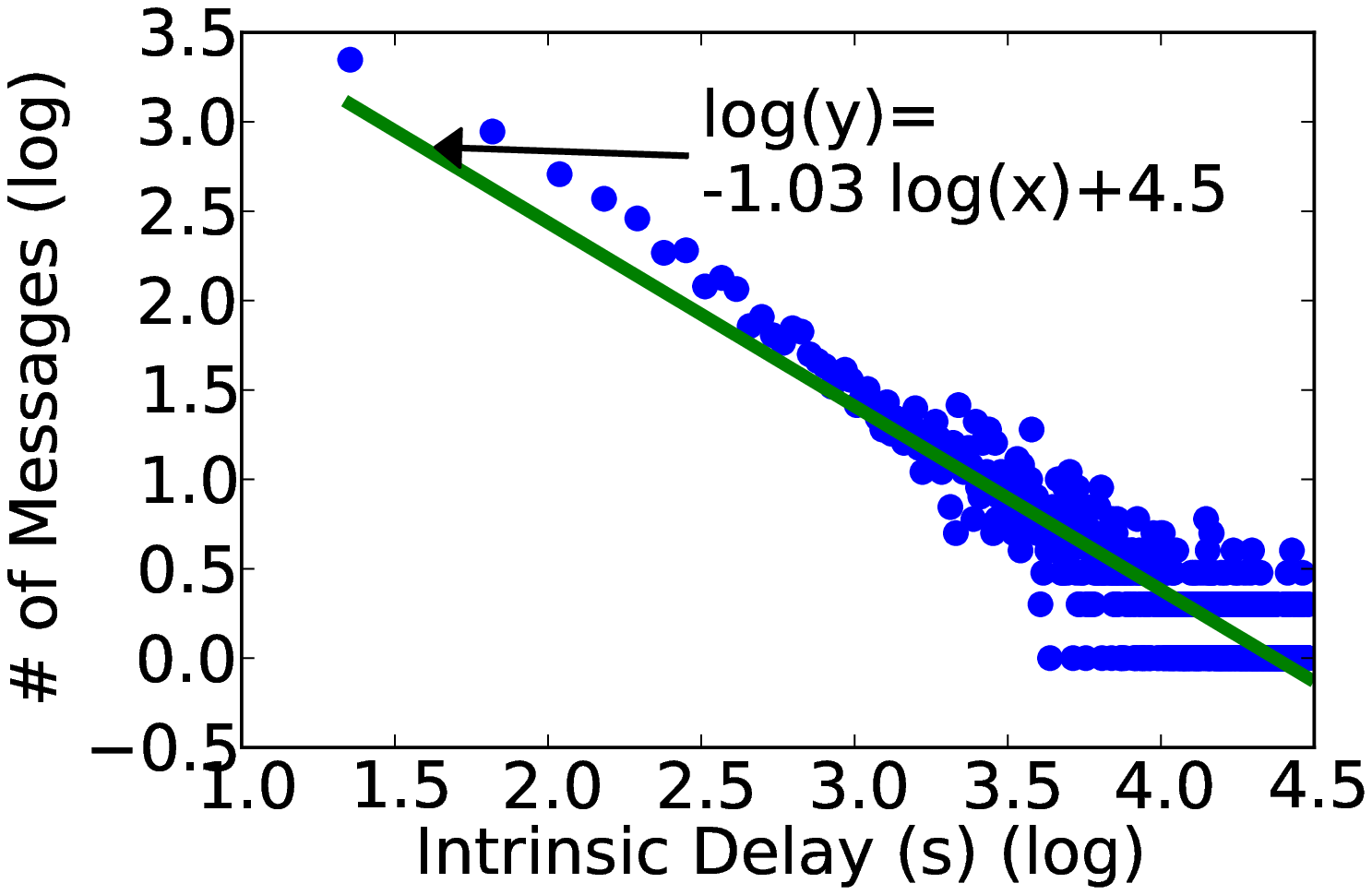}
        {\label{fig:distri_intrinsic_delay.eps}}
    }
    \subfigure[Forwarding Sequence Length]{
        \includegraphics[width=0.46\linewidth]{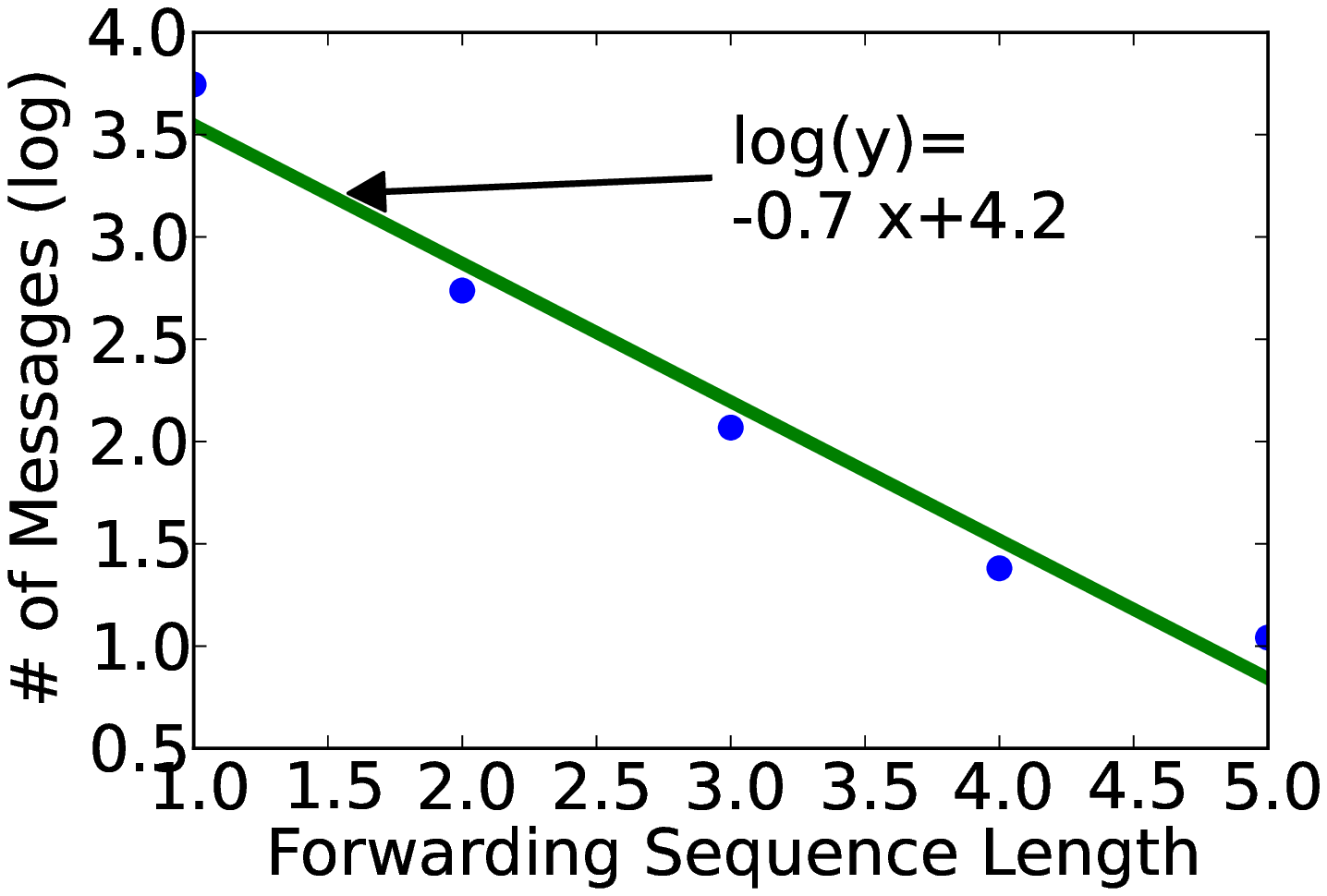}
        {\label{fig:distri_sequence_length.eps}}
    }
    \caption{Model Parameter Fitting}
\end{figure}

The empirical distribution of $I$ and $L$ are plotted in Figs.
\ref{fig:distri_intrinsic_delay.eps}-\ref{fig:distri_sequence_length.eps}. 
Observe that the intrinsic delay
of messages follows a power-law decay while exponential decay is exhibited by 
forwarding sequence length with their distributions given by:
\begin{equation}
p (i) = \frac{1}{Z_i} i^a 10^b \hspace{1.4em}, \hspace{2.4em} p (l) =
   \frac{1}{Z_l} 10^{cx + d} \nonumber
\end{equation}
where $a = - 1.03$, $b = 4.5$, $c = - 0.7$ and $d = 4.2$ based on fitting against experimental measurements. 
The normalization factors can be calculated as:
\begin{eqnarray}
    Z_i &=& \int_1^{30265} i^a 10^b \mathd i \nonumber\\
    &=& 10^b \left[ \frac{1}{a + 1}
   (i_{(\max)})^{a + 1} - \frac{1}{a + 1} (i_{(\min)})^{a + 1} \right]
   \nonumber
\end{eqnarray}
and
\[ Z_l = \int_1^{\infty} 10^{cl + d} \mathd l = \frac{- 1}{c \ln 10} e^{(c +
   d) \ln 10} \]
where $i_{(\max)} = 30265$ and $i_{(\min)} = 1$ are the maximum and minimum
intrinsic delay from observed from the trace data.

\subsubsection{Closed Form expression for Extra Forwarding Delay}

The expression for EFD, $\Epsilon [\Delta_{\tmop{EFD}}] = \Epsilon [L]
\Epsilon [\max \{ I, W \} - I]$, can be derived in two steps. Instead of going
through the fitted parameters, we estimate $\Epsilon [L]$ directly by
averaging over the forwarding sequence lengths and get $\Epsilon [L] = 1.14$.
This is to reduce possible errors in parameter fitting. The second part can be
solved by double integral:
\begin{eqnarray*}
  &  & \Epsilon [\max \{ I_k, W \} - I_k]\\
  & = & \int_1^{i_{(\max)}} \int_0^h p (i, w) [\max \{ i, w \} - i] \mathd w
  \mathd i\\
  & = & \frac{1}{h} \frac{1}{Z_i}10^b  [  \frac{1}{(a + 1) (a + 2) (a +
  3)} h^{a + 3} 
  - \frac{1}{2} h^2  \frac{1}{a + 1} 
  \\
  &&
  +
  \frac{h}{a + 2} -
  \frac{2-a(a+3)}{(a + 1) (a + 2) (a + 3)}
  ]
\end{eqnarray*}

\subsubsection{Extra Forwarding Delay Evaluation}

\begin{figure}[h]
    \centering
    {\includegraphics[width=0.6\linewidth]{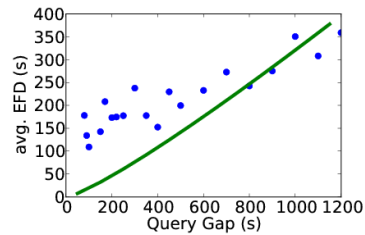}}
    \caption{Extra Forwarding Delay v.s. Query Gap\label{fig:h_vs_efd.eps}}
\end{figure}

Fig. \ref{fig:h_vs_efd.eps} depicts the scattered plots of the experiment results.
The analytical prediction is plotted as a solid line. One can see that the
analytical result matches experimental result when $h$ is relatively large ($>$ 10 mins). 
For smaller $h$, the analytical result under estimates the EFD which can be
explained by the following reasons: Firstly, we assume $h$ is the only delay but
there are other delays. Those small delays become dominating when $h$ is
small. Secondly, we assume the statuses are obtained immediately if one bot pulls
its neighbour. However, given the fully decentralized construction of our DSN,
one needs to pull RSS feeds from multiple friends. Since the network
performance between PlanetLab nodes varies widely, some bots may take much longer
time than QG to finish one round of pull. This also makes the experiment
results deviate from analytical ones with small $h$. One extreme case is when
$h = 0$, the analytical model, which assumes $h$ to be the only delay factor,
will predict EFD=0. This is obviously not the case and we are working on the
refinement of the model for EFD.

\subsubsection{Resource Consumption}

\begin{figure}[h]
    \subfigure[CPU]{
        {\includegraphics[width=0.46\linewidth]{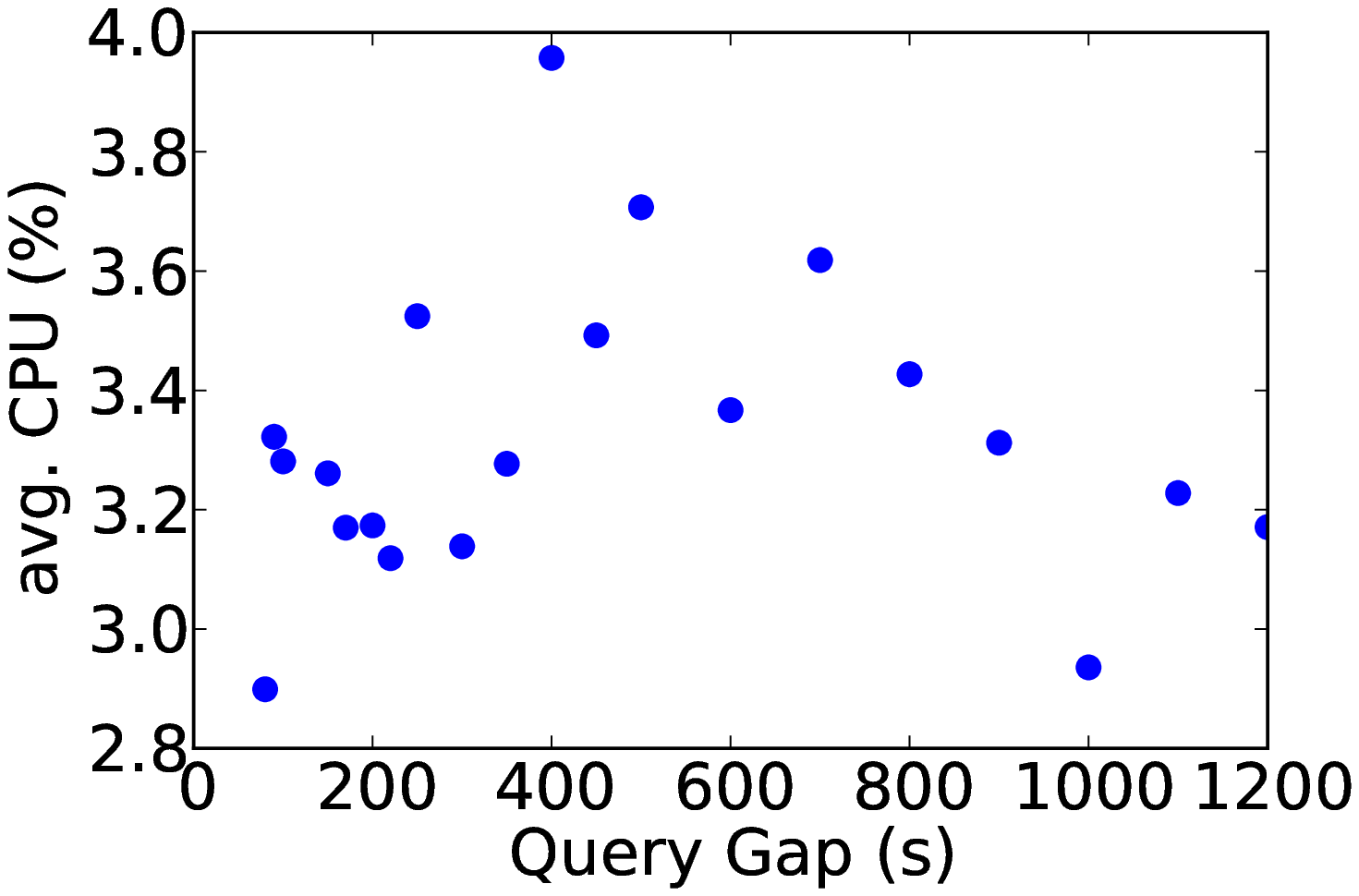}}
        {\label{fig:h_vs_cpu.eps}}
    }
    \subfigure[Memory]{
        {\includegraphics[width=0.46\linewidth]{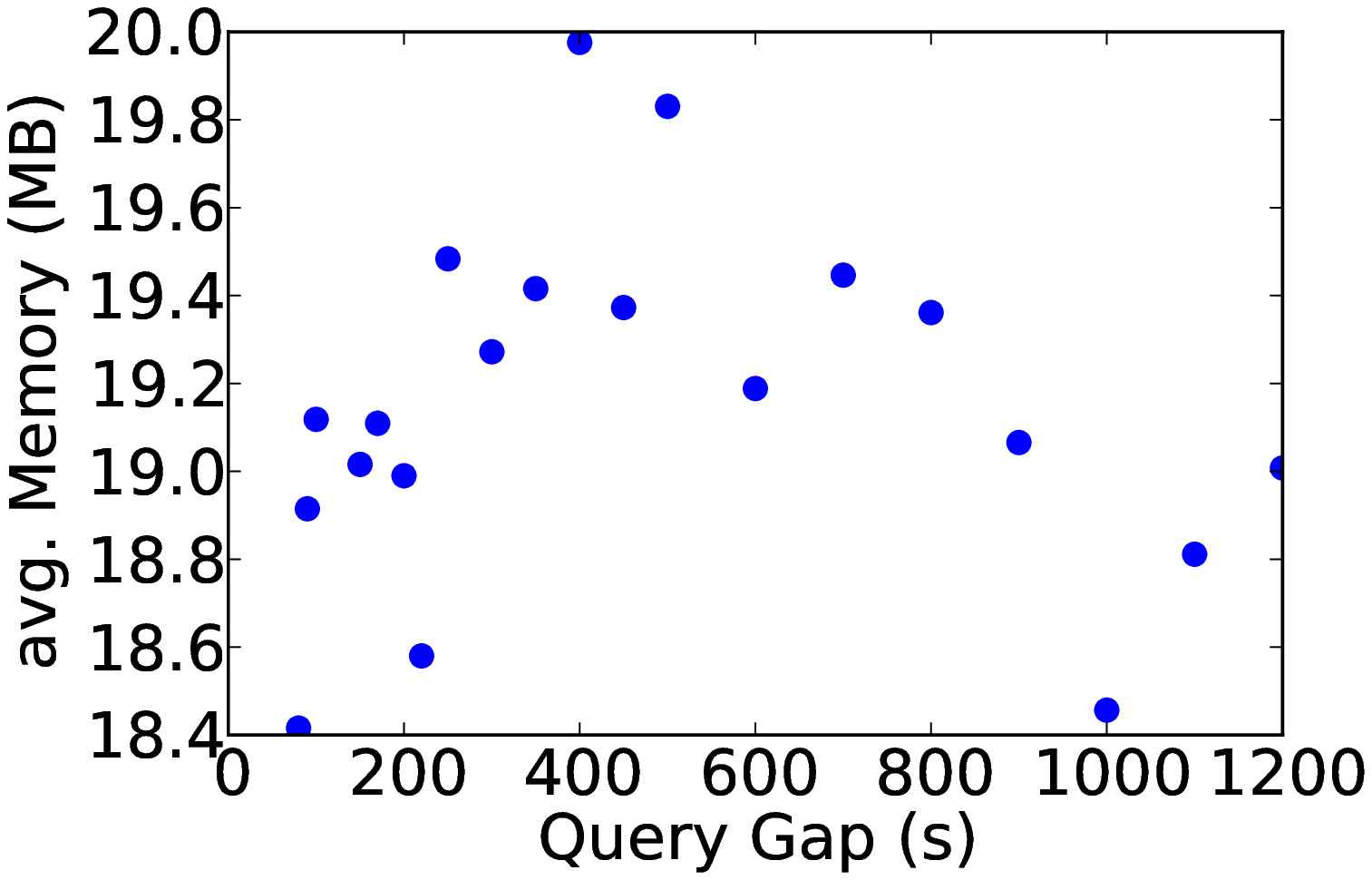}}
        {\label{fig:h_vs_mem.eps}}
    }
    \subfigure[HTTP Query]{
        {\includegraphics[width=0.46\linewidth]{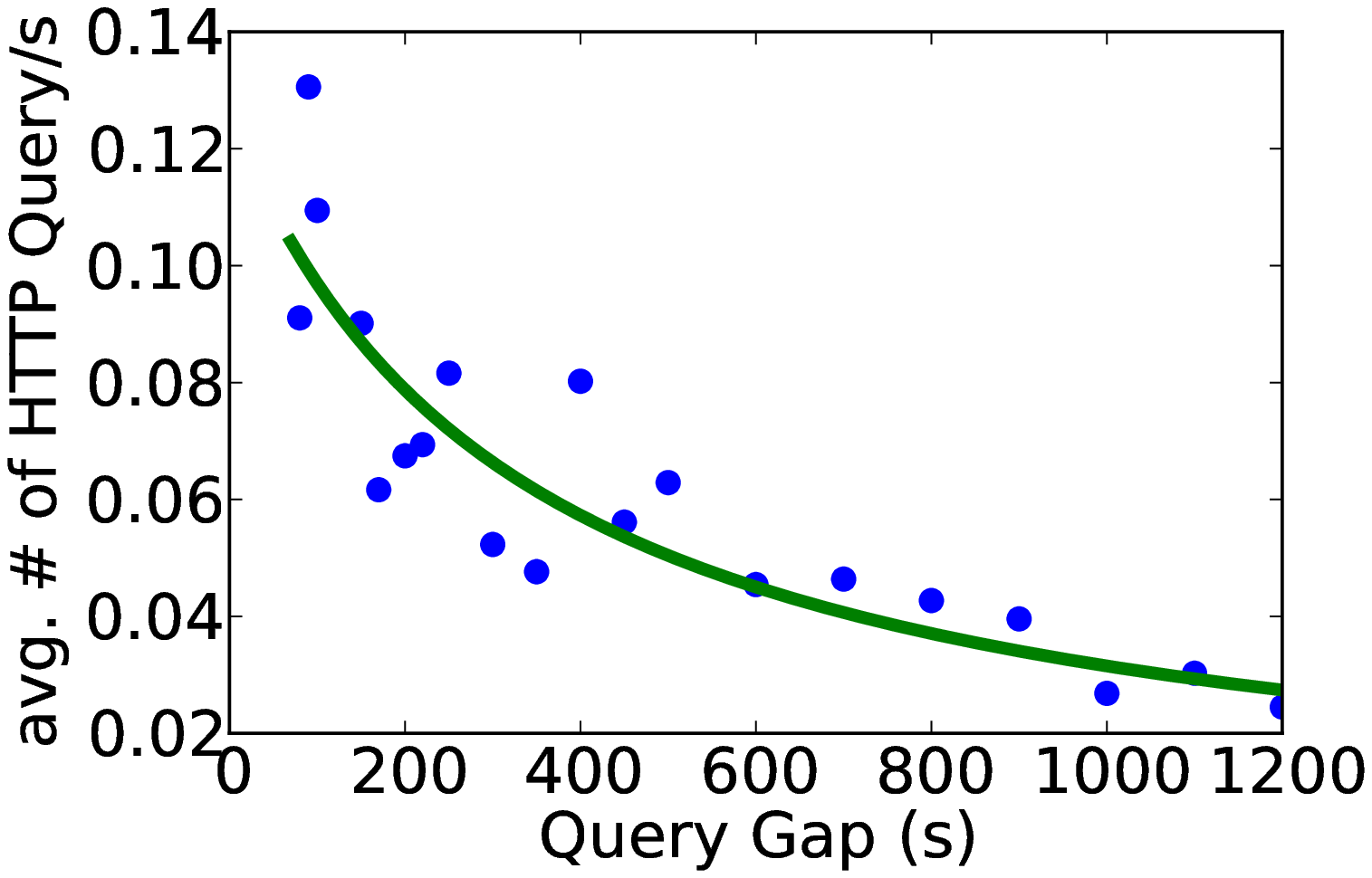}}
        {\label{fig:h_vs_http_query.eps}}
    }
    \subfigure[HTTP Size]{
        {\includegraphics[width=0.46\linewidth]{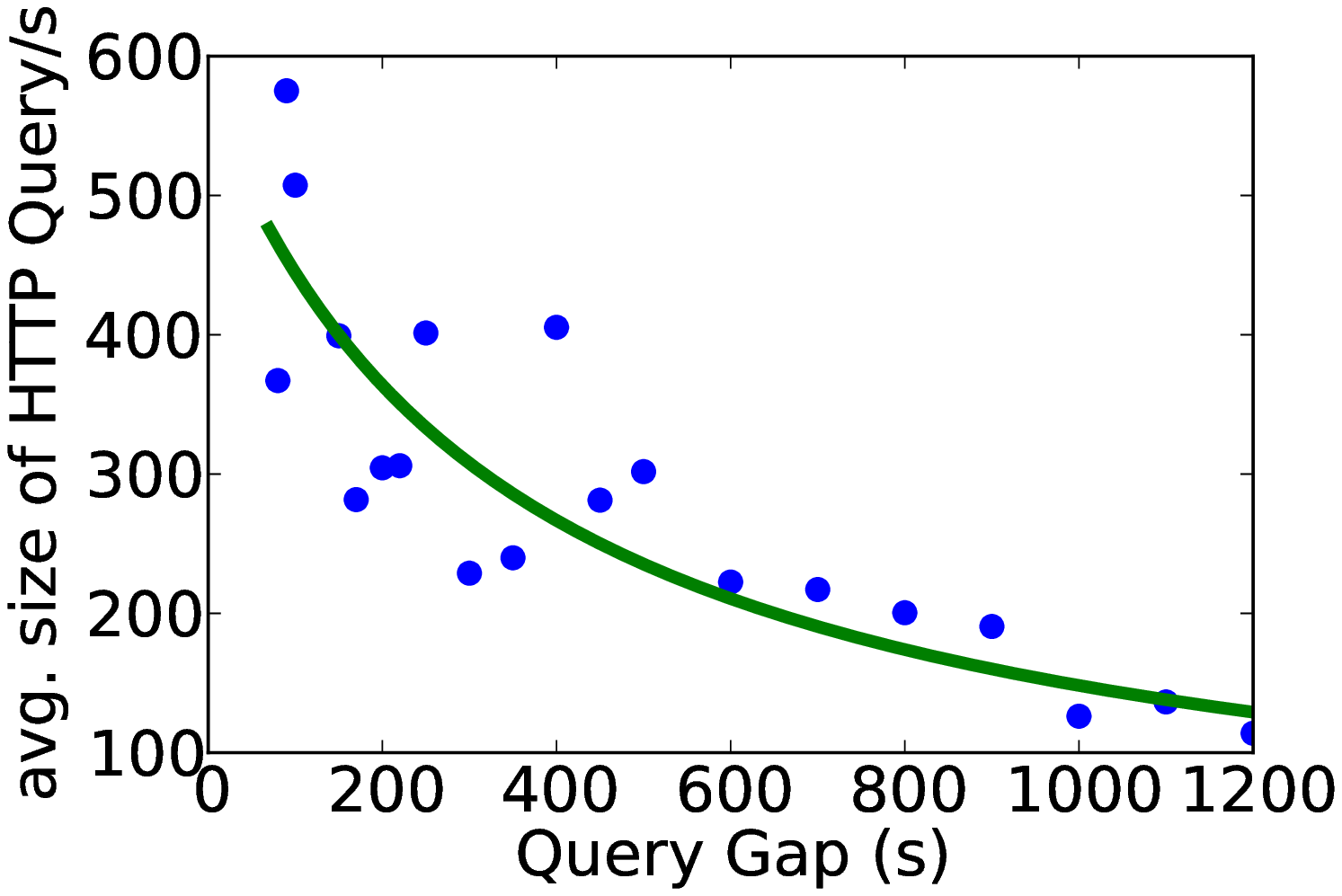}}
        {\label{fig:h_vs_http_size.eps}}
    }
    \caption{Consumption of Different Resources}
\end{figure}

Various resource consumption is depicted from Fig. \ref{fig:h_vs_cpu.eps} to Fig.
\ref{fig:h_vs_http_size.eps}. The results align well with our analysis in Section
\ref{sec:analytical-model-for-the-pull-based-backbone}, namely 1) CPU and
memory usage are just a constant plus experimental variance; 2) HTTP query and
size are reciprocal of Query Gap, $h$.

\section{Conclusions and Future Works}\label{sec:conclusions}

In this paper, we analyze the challenge towards a decentralized paradigm of
Social Networking Services and proposed a meta social networking approach to
solve the migration problem. We have built a cross-platform middleware, which is
lightweight, programmable, and flexible, to support the transition from
centralized OSNs to decentralized ones. We have presented the design and
implementation of SNSAPI and discussed design choices and observations based
on our year-long experience in maintaining and refactoring the corresponding
open-source project. We deployed a 6000-node DSN on PlanetLab based on SNSAPI 
to demonstrate that it is a viable approach to form a DSN. Using real
traces collected from a mainstream OSN, we show that with only mild resource
consumption, we can achieve acceptable forwarding latency comparable to that
of a centralized OSN. We have also developed an analytical model to
characterize the tradeoffs between resource consumption and message forwarding
delay in such a DSN. Via 20 parameterized experiments on PlanetLab, we have
found that the empirical measurement results match reasonably well with
performance predicted by our analytical model.

Future works include using non-uniform and even adaptive Query Gap, improving
the analytical model, formation of a DSN using push channels (e.g. Email
platform also supported by SNSAPI), developing a distributed back-off protocol
to alleviate the congestion in the DSN. During the deployment of the
6000-node DSN and the series of parameterized experiments, we have also built
a full set of tools for conducting and managing of similar large-scale DSN experiments. 
Like the SNSAPI and its Apps, this toolbox as well as the OSN activity traces we collected
from Sina Weibo will be open-sourced in the near future. 
We hope they will provide a foundation to support large-scale experimental deployment and
performance evaluation for DSN and the associated algorithms.

\end{document}